\documentclass[useAMS,usenatbib]{mn2e}
\input psfig.tex
\usepackage{latexsym}
\usepackage{natbib}
\usepackage{amssymb}
\usepackage{amsmath}
\title{Modelling galaxy spectra in presence of interstellar dust.\\
I. The model of ISM and the library of dusty SSPs}

\author[L. Piovan, R.Tantalo \& C. Chiosi]{Lorenzo Piovan$^{1,2}$, Rosaria Tantalo$^{1}$ \& Cesare Chiosi$^{1}$\\
 $^{1}$Department of Astronomy, University of Padova,
       Vicolo dell'Osservatorio 2, 35122 Padova, Italy\\
 $^{2}$Max-Planck-Institut f\"ur Astrophysik, Karl-Schwarzschild-Str. 1, Garching bei M\"unchen, Germany\\
E-mail: {\tt piovan@pd.astro.it; tantalo@pd.astro.it;
chiosi@pd.astro.it} }

\date{\tt Accepted: October 2005}

\pubyear{2005}

\begin{document}
\maketitle
\title{Galaxy spectra with interstellar dust}

\begin{abstract}

The advent of modern infrared astronomy has brought into evidence
the role played by the interstellar dust in galaxy
formation and evolution. Therefore, to fully exploit modern data,
realistic spectrophotometric models of galaxies must include this important
component of the interstellar medium (ISM).

In this paper, the first of a series of two devoted to modelling the
spectra of galaxies of different morphological type in presence of
dust, we present our description of the dust both in the
diffuse ISM and the  molecular clouds (MCs).

Our galaxy model contains three interacting components: the diffuse
ISM, made of gas and dust, the large complexes of MCs in which
active star formation occurs and, finally, the populations of stars
that are no longer embedded in the dusty environment of their
parental MCs.

Our model for the dust takes into account three components, i.e.
graphite, silicates and polycyclic aromatic hydrocarbons (PAHs). We
consider and adapt to our aims two prescriptions for the size
distribution of the dust grains and two models for the emission of
the dusty ISM. We cross-check the emission and extinction models of
the ISM by calculating the extinction curves and the emission for
the typical environments of the Milky Way (MW) and the Large and
Small Magellanic Clouds (LMC and SMC) and by comparing the results
with the observational data. The final model we have adopted is an
hybrid one which stems from combining the analysis of
\citet{Guhathakurta89} for the emission of graphite and silicates
and \citet{Puget85} for the PAH emission, and using the distribution
law of \citet{Weingartner01a} and the ionization model for PAHs of
\citet{Weingartner01b}.

We apply the model to calculate the spectral energy distribution
(SED) of Single Stellar Populations (SSPs) of different age and
chemical composition, which may be severely affected by dust at
least in two types of star: the young, massive stars while they are
still embedded in their parental MCs and the intermediate- and
low-mass AGB stars when they form their own dust shell around
\citep[see][for more details about AGB stars]{Piovan03}.

We use the "Ray Tracing" method to solve the problem of radiative
transfer and to  calculate extended libraries of SSP SEDs. Particular
care is paid to model the contribution from PAHs, introducing
different abundances of $C$ in the population of very small
carbonaceous grains (VSGs) and different ionization states in PAHs.
The SEDs of young SSPs are then compared with observational data of
star forming regions of four local galaxies  successfully
reproducing their SEDs  from the UV-optical regions to
the mid and far infrared (MIR and FIR, respectively).
\end{abstract}




\section{Introduction}\label{Introduction}

The interstellar dust, either tightly associated to
stars and/or dispersed in the ISM, has got more and more attention
over the years, in particular with the advent of modern infrared
satellites (e.g. IRAS, COBE and ISO), because of its role in many
astrophysical phenomena \citep[see][for more
details]{Draine03b,Draine03c,Draine03a,Draine04a}.

Leaving exceptions aside, there are at least three main
circumstances in which dust influences the stellar light: (i) It is
long known that for a certain fraction of their life very young
stars are embedded in the parental MCs. Even if the duration of this
obscured period is short, its effect  on the light emitted by these
stars cannot be neglected as a significant fraction of the light
(initially almost all) is shifted to the IR region of the spectrum.
(ii) Low and intermediate mass stars in the Asymptotic Giant Branch
(AGB) phase may form an outer dust-rich shell of material obscuring
and reprocessing the radiation emitted by the star underneath. (iii)
Finally, thanks to the contribution of metal-rich material by
supernovae and stellar winds in Wolf-Rayet and AGB stars, the ISM
acquires over the years a dust rich component. The UV-optical
radiation emitted by stars passing through this dust-rich
intergalactic gas is absorbed and then re-emitted in the far IR.

In this paper, first we develop a model for the absortiom/emission
properties of a dusty medium and second we apply it to derive the
SED of young stars still embedded in their parental MCs. In a
companion paper \citep{Piovan05}, we will present detailed
chemo-spectrophotometric models of galaxies of different
morphological type whose SED from the UV to the far IR is derived
including dusty MCs and the presence of diffuse, dust-rich ISM.

Stars are preferentially born inside massive, dense and cool MCs
characterized by low temperatures $\left(T\lesssim 9-15
\textrm{K}\right)$, masses in the range from $\sim 10^{4}$ to
$10^{6}M_{\odot}$ and dimensions from $\sim 6$ to $60pc$, the most
massive and big ones being less numerous \citep{Solomon87}.
Furthermore all regions with active star formation, e.g. in the MW,
are also associated to $H_{2}$  clouds, thanks to which the MCs are
mapped by means of radio surveys.  In general they seem to be
organized in hierarchical structures forming very complicate and
large complexes, inside which a large number of substructures of
higher density are found. The observation of the earliest
evolutionary phases of young stars is severely hampered by the
presence of dust, which absorbs and diffuses a large fraction of the
radiation emitted by the stars in the optical returning it in the
MIR/FIR. However, as soon as the first-born, massive stars evolve,
their strong stellar winds and mass-loss, intense ionizing radiation
fields and final explosion as type II supernovae will eventually
destroy the MCs in which they are embedded. Massive stars become
eventually visible in UV-optical regions of the spectrum. The time
scale for this to occur is $\sim 10^{6}-10^{8}$ years, i.e. the
lifetime of the most massive stars in the population. To summarize,
we are able to map the location of MCs thanks to radio data, to
follow the first evolutionary stages of star formation thanks to FIR
data and finally to directly observe the stars in the UV-optical
when the parental clouds have been evaporated. However all
intermediate stages are precluded to direct observations because
they are obscured by dust.  They must be inferred from theory, which
unfortunately is still far from being fully satisfactory.

The problem is particularly severe when the evolutionary population
synthesis technique (EPS) is applied to model the SED, the
integrated magnitudes and broad-band colors of galaxies, by folding
the properties of SSPs of different age and chemical composition on
the star formation history (SFH). Classical studies of this subject
\citep{Bruzual93,Bressan94,Tantalo96,Tantalo98a,Fioc97,Chiosi98}
ignore the presence of dust on the light emitted by  SSPs. In other
words, "bare" spectra of SSPs are used to synthesize the SED of a
galaxy. This approximation sounds acceptable when modelling the SED
of old systems, such as early-type galaxies, in which star formation
took place long ago, even if also in this case the presence of dust
around AGB stars should not be ignored  \citep{Piovan03,Temi04}.
This is certainly not the case with the late-type and starburst
galaxies that are rich of gas, dust and stars of any age. Therefore,
it is mandatory to include the effect of dust on the radiation
emitted by young stars. Because of the high density in the regions
of star formation, the optical depth may be very high also for IR
photons and the full problem of radiative transfer has to be
considered.

The plan of the paper is as follows. In Sect. \ref{dust_model} we
model in detail the extinction and emission properties of the dusty
ISM. In particular we present the optical properties of graphite,
silicates and PAHs, the two distribution laws for the grain sizes
(shortly named MRN ad WEI models), the cross sections and the dust
to gas ratio. In Sect. \ref{casting} we address the topic of
emission from dust grains, and present two models shortly indicated
as GDP and LID. The theory is applied to derive the average
extinction curves and emission properties for the diffuse ISM of the
MW, LMC and SMC (Sects. \ref{ext_emis}, \ref{ext_curves}, and
\ref{emission}, respectively). Then we apply our best solution for
the ISM extinction and emission to derive the SEDs of young dusty
SSPs. We summarize the fundamentals of spectral synthesis in Sect.
\ref{basic_popsyn} and in Sect. \ref{sp_dis_sgd} we present our
choice for the spatial distribution of young stars inside the MCs.
In Sect. \ref{rad_transport} we cast the problem of radiative
transfer introducing the key parameters and discussing the
dependence of the optical depth on the cloud physical parameters. In
addition to this, we quickly summarize  the the "ray tracing"
technique to solve the problem of radiative transfer throughout an
absorbing/emitting medium. In Sect. \ref{ised_dusty} we present  the
SEDs of very young SSPs surrounded by their MCs. Having in mind the
application of our results to studies of population synthesis in
galaxies \citep{Piovan05}, we have calculated a library of young
SEDs with dust for large ranges of the parameters, paying particular
attention to the emission of PAHs. In addition to this, we show
results for a simple way of modelling the evaporation of the MCs
surrounding young SSPs. In Sect. \ref{compare_observations} we
compare our SEDs with observational data of star forming regions. To
this aim, we considered the central regions of four starburst
galaxies, namely Arp$220$, NGC$253$, M$82$ and NGC$1808$. Finally,
in Sect. \ref{dis_concl} we draw some general remarks and
conclusions.

\section{Modelling the properties of dust}\label{dust_model}

Chief workbenches for the study of the physical properties of the
interstellar grains are the extinction curves and the IR emission of
dust observed with growing detail in different physical
environments. From the characteristic broad bump of the extinction
curves in the UV at $2175$ \AA \hspace{0.5mm} and the absorption
features at $9.7\mu m$ and $18\mu m$ \citep{Draine03a} we can
already infer that a two components model made of graphite and
silicates is required.  The study of the emission adds
further constraints. A
population of VSGs has been invoked to reproduce the emissions
observed by IRAS in the pass-bands $12\mu m$ and $25\mu m$. The VSGs
temperatures can fluctuate well above $20 \ \textrm{K}$ if their
energy content is small compared to the energy of the absorbed
photons \citep{Leger84,Desert86a,
Dwek86,Guhathakurta89,Siebenmorgen92,Draine01,Li01}. Excluding that
VSGs are made of silicates simply because the $10 \mu m$ emission
feature of silicates is not detected in diffuse clouds
\citep{Mattila96,Onaka96}, most likely they are made of carbonaceous
material with broad ranges of shapes, dimensions, and chemical
structures \citep[see also][for a discussion of this topic]{Desert86b,Li02c}.

Emission lines at $3.3$, $6.2$, $7.7$, $8.6$, and $11.3\mu m$,
originally named unidentified infrared bands (UIBs), have been
firstly observed in luminous reflection nebulae, planetary HII
regions and nebulae \citep{Sellgren83,Mathis90} and subsequently
also in the diffuse ISM with IRTS \citep{Onaka96,Tanaka96} and ISO
\citep{Mattila96}. There is nowadays the general consensus that
these lines owe their origin to the presence of PAH molecules,
vibrationally excited by the absorption of a UV-optical photon
\citep{Leger84,Li01}. Currently these spectral features are commonly
referred to as the aromatic IR bands (AIBs).

 Based on these considerations, any realistic model of a
dusty ISM, to be able to explain the UV-optical extinction and the
IR emission of galaxies, has to include at least three components,
i.e. graphite, silicates, and PAHs. Furthermore, while it may treat
the big grains as in thermal equilibrium with the radiation field,
it has to allow the VSGs to have temperatures above the mean
equilibrium value. In order to obtain the properties of a mixture of
grains, we have to specify their cross sections, their dimensions,
and the kind of interaction with the local radiation field.

\textbf{Optical properties.} The optical properties of PAHs,
silicate and graphite grains together with the corresponding
dimensionless scattering and absorption coefficients,
$Q_{sca}(a,\lambda)$ and $Q_{abs}(a,\lambda)$ have been taken from
\citet{Draine84}, \citet{Laor93}, \citet{Draine01} and \citet{Li01}.
These coefficients are defined as the ratio of the cross section
$\sigma $ to the geometrical section $\pi a^{2}$, where $a$ is the
dimension of the grain.

\textbf{Distribution laws: MRN-like and WEI models.} In the ISM the
grain dimensions are likely to vary over a large range of values.
Therefore in order to properly model the optical properties of the
ISM we need to specify the distribution law of the grain dimensions
and to fix their upper and lower limits. Two cases are considered.

In the first one, shortly referred to as "WEI", we adopt the
analytical law  proposed by  \citet{Weingartner01a}. It is
a very complicate relationship for the size distribution of
carbonaceous dust grains, which simultaneously deals with graphite
and PAHs and smoothly shifts from PAHs to small graphite grains,
considering PAHs as the extension of carbonaceous grains down
to the molecular regime:

\begin{eqnarray} \label{carbWein}
\frac{1}{n_{H}}\frac{dn_{g}}{da} &=&
D\left(a\right)+\frac{C_{g}}{a}\left(\frac{a}{a_{t,g}}\right)^{\alpha_{g}}
F \left(a, \beta_{g}, a_{t,g} \right) \times \nonumber \\
& \times & \negmedspace \left\{
\begin{array}{ll}
1 \qquad \qquad \qquad \qquad \hspace{3mm} 3.5 \textrm{\AA} < a < a_{t,g} \\
exp \left\{-\left[\left(a-a_{t,g}\right)/a_{c,g} \right]^{3}\right\}
\quad \hspace{1mm} a > a_{t,g} \\
\end{array}
\right.
\end{eqnarray}

\noindent where $C_{g}$ is the abundance of carbon, $a_{t,g}$ is a
transition radius securing a smooth cut-off for dimensions $a >
a_{t,g}$, $a_{c,g}$ is a parameter that controls the cut-off
steepness, and $\alpha_{g}$ is the exponent of the power law.
$D\left(a\right)$, the sum of two log-normal terms that helps to
better reproduce the emission by very small carbonaceous grains, is
given by

\begin{eqnarray}\label{log_normal_term}
\frac{1}{n_{H}}\left(\frac{dn_{g}}{da}\right)_{vsg} &=&
D\left(a\right) \nonumber \\
& = &\sum_{i=1}^{2}\frac{B_{i}}{a}
\exp\left\{-\frac{1}{2}\left[\frac{\ln\left(a/a_{0,i}\right)}{\sigma}\right]^{2}\right\}
\end{eqnarray}

\noindent for $a > 3.5$ \AA. Following \citet{Weingartner01a}, the
term $B_{i}$ is

\begin{eqnarray} \label{Bi_factor}
B_{i} & = & \frac{3}{\left(2 \pi \right)^{3/2}} \frac{\exp \left(
-4.5 \ \sigma^{2}\right)}{\rho a_{0,i}^{3} \sigma} \nonumber \\
& \times & \frac{b_{C,i}m_{C}}{1+erf\left[3 \sigma
\sqrt{2}+\ln\left(a_{0,i}/3.5 \AA\ \right)/\sigma \sqrt{2}\right]}
\end{eqnarray}

\noindent where $m_{C}$ is the mass of a C atom, $ \rho = 2.24 g \
cm^{-3}$ is the density of graphite, $ b_{C,1}= 0.75 \ b_{C}$,
$b_{C,2} = 0.25 \ b_{C}$ with $b_{C}$ the total C abundance (per H
nucleus) in the log-normal populations, $a_{0,1} = 3.5$ \AA,
$a_{0,2} = 30$ \AA\ and $\sigma = 0.4$. The function $F \left(a,
\beta_{g}, a_{t,g} \right)$ in eqn. $\ref{carbWein}$ is a suitable
correcting term of curvature [see eqn. (6) in \citet{Weingartner01a}
for more details]. Finally, to obtain the total abundance of carbon
we need to add the contributions coming from the two log-normal
distributions, i.e. $D\left(a\right)$, to the term controlled by
$C_{g}$. An interesting property of eqn. (\ref{carbWein}) is that
thanks to its ad-hoc analytical form very good fits of the
extinction curves are possible by varying the contribution of the
very small carbonaceous grains to the $C$ abundance in the ISM, a
quantity not yet firmly assessed. As shown by
\citet{Weingartner01a}, the sole extinction curve cannot constrain
the carbon abundance $b_{C}$, but only provide an upper limit. For
any ratio of visual extinction to reddening
$R_{V}=A\left(V\right)/E\left(B-V\right)$, the limit is reached when
the very small carbonaceous grains (PAHs and small graphite grains)
are able to account for all the ultraviolet bump of the extinction
curve. A relationship similar to eqn. (\ref{carbWein}) is adopted in
\citet{Weingartner01a} for the silicate grains but, since there is
no need for log-normal distributions of VSGs, the abundance of
silicates is completely described by the parameter $C_{s}$. With the
aid of the above relations by \citet{Weingartner01a} it is possible
to reproduce the average extinction curves of the diffuse and dense
ISM of the MW, LMC and SMC.

In the second case, shortly referred to as "MRN", we have modified
and extended the power-law  proposed by \citet{Mathis77}. To better
explore all the possible situations, we split the distribution law
of the i-th component in several intervals labelled by $k$

\begin{equation}\label{distribnew}
\frac{1}{n_{H}}\frac{dn_{i}}{da}=\left\{
\begin{array}{ll}
    A_{i}a^{\beta+\beta_{1}}a_{b_{1}}^{-\beta_{1}}, & a_{c_{1}}<a<a_{c_{2}} \\
    A_{i}a^{\beta+\beta_{2}}a_{b_{2}}^{-\beta_{2}}, & a_{c_{2}}<a<a_{c_{3}} \\
    ...................., & ..................... \\
    A_{i}a^{\beta+\beta_{k-1}}a_{b_{k-1}}^{-\beta_{k-1}}, & a_{c_{k-1}}<a<a_{c_{k}} \\
\end{array}
\right.
\end{equation}

\noindent with the condition that $
a^{\beta+\beta_{j}}a_{b_{j}}^{-\beta_{j}}\approx
a^{\beta+\beta_{j+1}}a_{b_{j+1}}^{-\beta_{j+1}}$
 where $j=1,...,k-2$ if $k>2$. The meaning of
$a_{c_{k}}$ is obvious, whereas the $a_{b_{k}}$ are the "modulation
factors" introduced by \citet{Draine85}. For $k=2$ relation
(\ref{distribnew}) reduces to  a single power-law distribution with
$a_{c_{1}}=a_{min}$ and $a_{c_{k}}=a_{max}$. Usually,  $k=2$ for
PAHs, and $2 \leq k \leq 4$ for silicates and carbonaceous grains.
In most cases we  adopt $k=2$ or $k=3$. More complicated power laws
are obtained with $k=4$.

\textbf{Cross sections} The cross sections of scattering, absorption
and extinction for the grain mixture in each component of the ISM
are defined as

\begin{equation} \label{sigabs}
\sigma_{p,i}\left( \lambda \right)
=\int_{a_{\min,i}}^{a_{\max,i}}\pi a^{2}Q_{p,i}\left( a,\lambda
\right) \frac{1}{n_{H}}\frac{dn_{i}(a)}{da} da
\end{equation}

\noindent where the index $p$ stands for absorption (abs),
scattering (sca), total extinction (ext) and the index $i$
identifies the type of grains, $a_{min,i}$ and $a_{max,i}$ are the
lower and upper limits of the size distribution for the i-type of
grain, and finally $n_{H}$ is the number density of $H$ atoms. The
total dimension-less extinction coefficient is $Q_{ext}\left(
a,\lambda \right)= Q_{abs}\left(a,\lambda
\right)+Q_{sca}\left(a,\lambda \right)$.

With the aid of the above cross sections it is possible to calculate
the optical depth $\tau_{p}(\lambda)$ along a given path

\begin{equation}
\tau_{p}\left( \lambda \right) =\sum_{i} \sigma _{p,i}\left( \lambda
\right) \int_{L}n_{H}dl=\sum_{i} \sigma_{p,i}\left(\lambda \right)
\times N_{H} \label{tauabs}
\end{equation}

\noindent where $L$ is the optical path and the meaning of the other
symbols is the same as before. In this expression for the various
$\tau_{p}(\lambda)$ we have implicitly assumed that the cross
sections remain constant along the optical path.

\textbf{Dust to gas ratio} The coefficients $b_{C}$, $C_{g}$,
$C_{s}$ for the WEI model and $A_{i}$ for the MRN model fix the
abundances of carbon and silicates with respect to the abundance of
hydrogen. In order to apply our models to a wide range of
metallicities and ages, it is important to discuss how these
coefficients would change as function of the chemical composition of
the ISM (the metallicity at least). The dust-to-gas ratio is defined
as $\delta =M_{d}/ M_{H}$  where $M_d$ and $M_{H}$ are the total
dust and hydrogen mass, respectively. If $m_{i}$ is the total mass
of the $i$-th type of grains then $M_{d}=\sum_{i}m_{i}$, where the
mass $m_i$ is obtained from integrating over the grain size
distribution (which depends also on the abundance coefficients).
Knowing the gas content of a galaxy (in principle function of the
age), the amount of dust in the ISM depends on $\delta$. For the MW
and other galaxies of the Local Group, $\delta$ is estimated to vary
from $1/100$ to $1/500$. In models for the MW, $\delta = 0.01$ is
often adopted, while $\delta =0.00288$ and $\delta =0.00184$ are
typical values for the LMC and SMC. The mass-density ratios of
interstellar dust are roughly in the proportion  $1:1/3:1/5$ going
from the MW to the LMC and SMC \citep{Pei92}. A simple way of
incorporating the effect of metallicity is to assume that $\delta
\varpropto Z$ in such a way to match the approximate average results
for MW, LMC and SMC: $\delta
=\delta_{\odot}\left(Z/Z_{\odot}\right)$. \noindent This relation
simply implies that metal-rich galaxies are also dust-rich. The
above relation $\delta \varpropto Z$ agrees with the results by
\citet{Dwek98} based on evolutionary models for the compositions and
abundances of gas and dust in the MW. However, it oversimplifies the
real situation.

The slope of the extinction curves greatly varies passing from the
MW to  LMC and SMC.  Moreover, the bump due to graphite decreases
from MW to LMC and SMC \citep{Calzetti94}. These differences have
been attributed to the metallicity decreasing from $Z=Z_{\odot }$ in
the solar vicinity, to $Z=\frac{1}{3}Z_{\odot }$ in  the LMC, and
$Z=\frac{1}{5}Z_{\odot}$ in the SMC. However, the metallicity
difference does not only imply a difference in the absolute
abundance of heavy elements in the dust, but also a difference in
the relative proportions, i.e. in the composition pattern. Probably
the ratio graphite to silicates varies from galaxy to galaxy.
Despite these uncertainties \citep{Devriendt99}, the relation
$\delta \varpropto Z$ is adopted to evaluate the amount of dust in
galaxy models \citep[e.g.][]{Silva98} by simply scaling the dust
content adopted for the ISM of the MW to the metallicity under
consideration. In this study, using the data for the extinction
curves and emission of the MW, LMC and SMC, we seek to build  a
model of the dusty ISM describing the effect of different
metallicities in a more realistic way. It is clear, however, that
the problem remains unsettled for metallicities higher than the
solar one, for which we have no information. The relation $\delta
\varpropto Z$ together with the relative proportions holding good
for  the MW extinction curve, can be reasonably adopted also for
other dusty ISMs characterized by metallicities close to the one
of the MW.

\section{Emission models}\label{casting}

In the following, two cases are considered to evaluate the emission
of a dusty ISM. First, the easy to calculate model that stems from
the thermal-continuous approach by \citet{Guhathakurta89} for the
temperature fluctuations of silicate and graphite grains and a
modification of the prescriptions by \citet{Puget85} for the
temperature fluctuations of PAHs. The calculation of the ionization
state of PAHs is also properly introduced following
\citet{Weingartner01a}. We will refer to it as the "GDP" model.
Second, a more elaborated model in which we implement the
state-of-the art for the temperature fluctuations of the very small
grains as recently proposed by \citet{Draine01} and \citet{Li01}.
This model will be shortly indicated as the "LID" model.

\subsection{The GDP model} \label{GDP}

\textbf{Emission from silicate and graphite grains.} In order to
calculate the temperature distribution acquired by small spherical
grains of dust heated by photons and collisions with energetic
particles, it is useful to define a vector state $P\left( t\right)$.
The state of a grain is determined by its enthalpy $E$ and the
components $P_{k}\left( t\right)$ of the state vector are the
probability for a grain to find itself in the $k$-th bin of enthalpy
at the time $t$. The enthalpy spans the range $\left[ E_{\min},
E_{\max}\right ] $ which in turn  splits into $N$ discrete bins each
of which characterized by $E_{k}$ and $\Delta E_{k}$, i.e. the
average value of the enthalpy  and the width of the enthalpy bin. The
enthalpy grid we have adopted is the same as in
\citet{Guhathakurta89}.

Let us also define the transition matrix
$A_{f,i}$ representing the probability for unit time that a grain in
the state $i$ may undergo a transition to the state $f\neq i$.
Neglecting the transitions to enthalpies outside the interval
$\left[ E_{\min },E_{\max}\right]$, the condition of statistical
equilibrium (steady state) is given by $dP_{f}^{ss}/dt = 0$
\citep{Guhathakurta89,Draine01}. Taking into account all the
transitions to and from any  bin we have
$\sum\limits_{i=1}^{N}A_{f,i}P_{i}^{ss}=0$. Together with the
normalization $\sum_{i=1}^{N}P_{i}^{ss}=1$ condition, it forms a system of linear
equations whose solution yields the equilibrium temperature
distribution. All technical details on the solution of this steady
state system  and the transition matrix $A_{f,i}$ can be found in \citet{Guhathakurta89} and
\citet{Draine01}.

The enthalpy and temperature intervals are usually chosen in such a
way that the distribution function $P_{k}\left( T\right)$ smoothly
changes from one bin to another and  the energy balance between
absorption and emission is conserved. \citet{Lis91} use a grid with
$N\simeq 200$. However \citet{Siebenmorgen92} note that grids with
high $N$ do not necessarily improve the accuracy of
$P_{k}\left(T\right)/\Delta E_{k}$ and accordingly adopt $N\simeq
60$, even if $N\simeq 30$ could be sufficient in some cases. In the
latest models of \citet{Draine01} the number $N$ of bins is in the
range $300 < N < 1500$, each bin (the first ones in particular)
contains at least two vibrational states, the bin widths smoothly
vary and the last bin is chosen at high enough energy so that the
expected population is very small (probability $P_{N} < 10^{-14}$).
Here we adopt $N\simeq 150-200$, which seems to secure computational
speed and accuracy at the same time.

A question soon arises: below which dimensions $a_{flu}$  must the
temperature fluctuations  be taken into account? According to
\citet{Guhathakurta89} and \citet{Siebenmorgen92}, the typical
dimension separating small from large grains is  about
$a_{flu}\simeq 150$\AA\, independently of the radiation field.
However, this conclusion was questioned by \citet{Li01} who found
that also for dimensions of the order of $a_{flu}\simeq 200$\AA\ the
temperature distribution is sufficiently broad thus introducing some
uncertainty in the estimate of the emission. To take this remark
into account we adopt here $a_{flu}\simeq 250$\AA. Finally, the
emission $j_{\lambda }^{small}$ for small graphite and silicate
grains is given by

\begin{eqnarray} \label{smallemission}
j_{\lambda }^{small}&=&\pi \int\nolimits_{a_{min
}}^{a_{flu}}\int\nolimits_{T_{min}}^{T_{max}} a^{2}Q_{abs}\left( a,
\lambda \right)B_{\lambda }\left( T\left(
a\right) \right)\nonumber \times \\
&& \times \frac{dP\left(a\right)}{dT} dT
\frac{1}{n_{H}}\frac{dn\left(a\right)}{da}da
\end{eqnarray}

\noindent where $dP \left(a\right)/dT $ is the distribution
temperature obtained for the generic dimension $a$, and $B_{\lambda
}\left( T\left( a\right) \right) $ is the Planck function.

In the case of grains with dimensions larger than about $\sim
250\textrm{\AA}$ which remain cool because of their high thermal
capacity and re-emit the energy absorbed from the incident radiation
field into the FIR, a good approximation is to evaluate the
equilibrium temperature $T\left(a\right)$ from the balance between
absorption and emission

\begin{equation}
\int Q_{abs}\left( a,\lambda \right) J_{\lambda }d\lambda =\int
Q_{abs}\left( a,\lambda \right) B_{\lambda }\left( T\left( a\right)
\right) d\lambda
\end{equation}

\noindent where $J_{\lambda}$ is the incident radiation field. With
the aid of  $T\left(a\right)$ we calculate the emission assuming
that big grains of fixed composition and radius behave like a black
body. Therefore, the emission coefficient is

\begin{equation} \label{bigemission}
j_{\lambda}^{big}=\pi \int_{a_{flu}}^{a_{max}}  a^{2} Q_{abs} \left(
a,\lambda \right) B_{\lambda} \left( T \left(a \right) \right)
\frac{1}{n_{H}}\frac{dn\left( a\right)}{da} da
\end{equation}

\textbf{Emission from PAHs.} Let us consider a photon of energy
$E=h\nu ^{^{\prime }}=hc/\lambda^{^{\prime }}$ impinging on a PAH
molecule. The maximum temperature increase caused by the absorption
of the
 photon is given by the solution of the equation

\begin{equation}
\frac{hc}{\lambda ^{^{\prime
}}}=\int\nolimits_{T_{min}}^{T_{max}\left( \lambda ^{^{\prime
}},a\right) }C_{PAH}\left( T,a\right) dT
\end{equation}

\noindent where $T_{min}\thicksim 5 \ \textrm{K}$ and
$C_{PAH}\left(T,a\right)$ is the thermal capacity of a grain of
dimension $a$ at the temperature $T$. As pointed out by
\citet{Xu89}, $T_{max}(\lambda ^{^{\prime}},a)$ is essentially
independent from the choice of $T_{min}$. In order to calculate
$C_{PAH}\left(T, a\right) $, we use the numerical fit of the
\citet{Leger89a} relationship proposed by \citet{Silva98}:

\begin{equation}
\frac{C_{PAH}\left( T,a\right) }{C_{max}}=\left\{
\begin{array}{llll}
mT    & & & T\leq 800 \ \textrm{K}   \\
nT+p  & & & 800 \ \textrm{K}\leq T\leq 2100 \ \textrm{K}   \\
1     & & & T\geq 2100 \ \textrm{K}
\end{array}
\right.
\end{equation}

\noindent where $m = 9.25\cdot 10^{-4}$, $n = 2\cdot 10^{-4}$, $p =
0.58$, and $C_{max}=3\left[ \left(N_{C}+N_{H}\right)-2\right ]
k_{B}$ \citep{Leger89a}. $N_{C}$ and $N_{H}$ are the number of
hydrogen and carbon atoms in the molecule, and $k_{B}$ is the
Boltzmann constant.

Dealing  with the number $N_{H}$ of H atoms in a PAH an important
issue has to be clarified. The PAHs in the ISM are subject to many
physical and chemical processes, chief among which are the
ionization and photo-dissociation caused by the absorption of a UV
photon, electron recombination with PAH cations, and chemical
reactions  between  PAHs and other major elemental species in the
ISM.

PAHs can be dehydrogenated (some of the $C-H$ bonds can be broken),
fully hydrogenated (one H atom for every available C site) or highly
hydrogenated (some of the peripheral $C$ atoms bear two $H$ atoms).
The reason is that  $H$ atoms are less strongly bound ($4.5eV$) than
carbon atoms ($7.2eV$) and therefore can be more easily removed by
energetic photons \citep{Leger89b}.

Over the years many studies have been devoted to model the
hydrogenation state at various levels of sophistication
\citep{Allain96b,Allain96a,Vuong00} and recently \citet{LePage01}
and \citet{LePage03} presented a very detailed model for the charge
and hydrogenation state of PAHs. The number of $H$ atoms is given by
the relation $N_{H}=\chi_{H} \cdot N_{S}$ where $N_{S}$ is the
number of available sites for hydrogen to bind to carbon and
$\chi_{H}$ is the hydrogenation factor. In a PAH with full hydrogen
coverage, the bond between $N_{S}$ and $N_{H}$ depends on nesting
the hexagonal cycles into the molecule. These cycles can form (i) an
elongated low-density object, named catacondensed PAH (linear chains
of n-cycles with  general scheme $C_{4n+2}H_{2n+4}$, like
naphthalene with 2 cycles, anthracene with 3 cycles, and pentacene
with 5 cycles); (ii) a compact object named pericondensed PAH
described by the formula $C_{6p^{2}}H_{6p}$, like coronene with
$p=2$, ovalene and circumcoronene or (iii) an untied object
(biphenile). The more compact the PAH, the more stable it is.

In agreement with this, to represent the main characteristics of the
interstellar PAHs, \citet{Omont86}, \citet{Dartois97} and
\citet{Li01} consider models of pericondensed PAHs, coronene-like,
that are more stable against photodissociation by UV photons.

From the formula for compact pericondensed PAHs with full hydrogen
coverage, $C_{6p^{2}}H_{6p}$, we get $N_{S}=6p=\left( 6\cdot
6p^{2}\right)^{0.5}=\left( 6N_{C}\right)^{0.5}$. As the number
$N_{C}$ of carbon atoms is known, we can derive the number $N_{S}$
of sites available for bonds. For example, for the coronene with
$p=2$ and $N_{C}=24$, we have $N_{S}=12$; if the coverage is
full with $\chi_{H}=1$, we have $N_{H}=12$. We are left with the
hydrogenation $\chi_{H}$ factor to be determined. This will depend
on the ISM conditions. For typical astronomical environments,
\citet{Siebenmorgen92} have suggested dehydrogenated values from
$0.01$ to $0.5$, without excluding values as low as $0.01$. More
recent studies by \citet{LePage01} and \citet{LePage03} adopt
hydrogenation factors that depend on the PAH dimensions. Their
analysis splits the population of PAHs in several groups. At the
typical conditions of the diffuse ISM, their general conclusion is
that very small PAHs (those with less than $30$ $C$ atoms) are
extremely dehydrogenated, larger PAHs with more than $30$ $C$ atoms
have full hydrogen coverage, whereas very large PAHs (with more than
150-200 $C$ atoms) can be highly hydrogenated. As the model
proposed by \citet{LePage01} and \citet{LePage03} is too much
detailed for the purposes of this study, the $H/C$ ratios in usage
here are taken from \citet{Li01}

\begin{equation}\label{coverage}
H/C=\left\{
\begin{array}{ll}
    0.5                   &\quad   N_{C}\leq 25    \\
    0.5/\sqrt{N_{C}/25}   &\quad   25 \leq N_{C} \leq 100  \\
    0.25                  &\quad   N_{C} \geq 100
\end{array}
\right.
\end{equation}

\noindent They yield $N_{H} = N_{S} = \left( 6N_{C}\right) ^{0.5}$
in the range $25 \leq N_{C} \leq 100$, which is typical of a
pericondensed PAH with full hydrogen coverage, while only for
smaller PAHs there is a partial dehydrogenation.

To calculate the emission of the PAHs we start defining the power
irradiated by a molecule of dimension $a$ at the temperature $T$

\begin{equation}
F\left( T,a\right) =\int_{0}^{\infty }\pi a^{2}Q_{abs}^{PAH}\left(
a,\lambda \right) \pi B\left( \lambda ,T\right) d\lambda
\end{equation}

\noindent where $B\left(\lambda,T\right) $ is the Planck function.
Once heated to a temperature $T$, the molecule cools down at the
rate

\begin{equation}
\frac{dT}{dt}=\frac{F\left( T,a\right) }{C_{P}\left( T,a\right) }
\end{equation}

\noindent The energy emitted at the wavelength $\lambda$ by a PAH of
dimension $a$, as a consequence of  absorbing a single photon with
energy  $hc/\lambda ^{^{\prime }}$, is given by

\begin{eqnarray}
S &=&\int\nolimits_{0}^{t}\pi a^{2}Q_{abs}^{PAH}\left( a,\lambda
\right) \pi B\left( \lambda ,T\right) dt=
\nonumber \\
&=&\int\nolimits_{T_{min}}^{T_{max}}\pi a^{2}Q_{abs}^{PAH}\left(
a,\lambda \right) \pi B\left( \lambda,T\right) \frac{C_{P}\left(
T,a\right) }{F\left(T,a\right) }dT
\end{eqnarray}

\noindent where $S = S(\lambda ^{^{\prime }},\lambda ,a)$ and
$T_{max} = T_{max}(\lambda ^{^{\prime }},a)$.

Under the effect of an incident radiation field $I(\lambda
^{^{\prime }})$ containing photons of any energy, the emission at
the wavelength $\lambda$ per $H$ atom becomes

\begin{eqnarray} \label{PAHemission}
J\left( \lambda \right)
&=&\frac{1}{n_{H}}\int\nolimits_{a_{PAH}^{low}}^{a_{PAH}^{high}}
\frac{dn_{PAH}\left(a\right)}{da}\pi
a^{2}Q_{abs}^{PAH}\left( a,\lambda \right) \times \nonumber \\
&& \times \int\nolimits_{0}^{\infty }\frac{I( \lambda ^{^{\prime }})
}{hc} \lambda ^{^{\prime }}\pi a^{2}Q_{abs}^{PAH}\left( a,\lambda
^{^{\prime}}\right) \times \nonumber \\
&& \times \int\nolimits_{T_{min}}^{T_{max}}\pi B\left( \lambda
,T\right) \frac{C_{P}\left( T,a\right) }{ F\left( T,a\right) }dT \,
d\lambda^{^{\prime }}\, da
\end{eqnarray}

\noindent from which

\begin{eqnarray} \label{fluxPAH}
J\left( \lambda \right) &=&\frac{\pi}{n_{H}hc}\int\nolimits_{\lambda
_{MIN}}^{\lambda _{MAX}}I\left( \lambda ^{^{\prime }}\right) \lambda
^{^{\prime
}}\int\nolimits_{a_{PAH}^{low}}^{a_{PAH}^{high}}\frac{dn_{PAH}\left(a\right)}{da} \nonumber \\
&&a^{2}Q_{abs}^{PAH}\left( a,\lambda^{^{\prime }}\right) S\left(
\lambda^{^{\prime}},\lambda ,a\right) da\, d \lambda^{^{\prime}}
\end{eqnarray}

\begin{figure}
\psfig{file=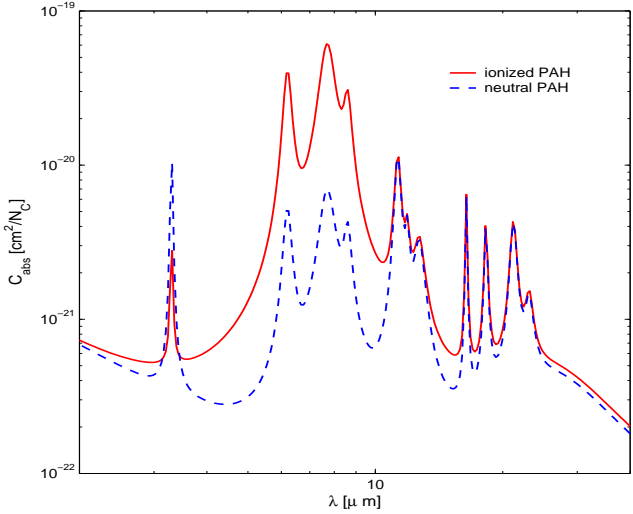,width=8.5truecm,height=6.9truecm}
\caption{Optical properties of neutral and ionized PAHs containing
$40$ atoms of carbon in the IR spectral region of the UIBs. The
difference between neutral (dashed line) and ionized PAHs is soon
evident: the ionization enhances the stretching modes C-C ($6,2$ and
$7.7\,\mu m$) and the bending mode C-H ($8.6\mu m$), whereas  it
weakens the stretching mode C-H at $3.3\,\mu m$ \citep{Li01}.}
\label{PAH2}
\end{figure}

The many  types of PAHs can differ not only for the hydrogen
coverage on peripheral carbon atoms, but also for the ionization
state. Because of their low ionization potential (about $6-7 \ eV$),
on one hand the PAHs can be ionized by  photo-electric effect
\citep{Bakes94}, on the other hand they can also acquire charge
through collisions with ions and electrons \citep{Draine87}.
Consequently, there is always some probability that a PAH is in a
charged state. Therefore, \citet{Li01} have calculated and made
available PAHs cross sections both for ionized and neutral PAHs.
Since the ionization state will affect the intensity of  AIBs, as we
can see in Fig. \ref{PAH2}, in our calculations we have included the
ionization of PAHs.

To this aim, the charge-state distribution model of PAHs is taken
from \citet{Weingartner01b}, who improved upon previous attempts by
\citet{Draine87} and \citet{Bakes94}. Indicating with $Z$ the charge
state of a PAH molecule, in conditions of statistical equilibrium we
have the following relation

\begin{equation}\label{statistical}
f\left(Z\right)\left[J_{pe}\left(Z\right)+J_{ion}\left(Z\right)\right]=
f\left(Z+1\right)J_{e}\left(Z+1\right)
\end{equation}

\noindent where $f\left(Z\right)$ is the probability for a PAH  to
possess the charge $Z$ (either positive or negative), $J_{pe}$ is
the photo-emission rate, $J_{ion}$ is the positive ion accretion
rate, and finally $J_{e}$ is the electron accretion rate. Following
\citet{Draine87}, $f\left(Z\right)$ is derived by recursively
applying  eqn. (\ref{statistical})
 to $f\left(0\right)$. We  obtain

\begin{equation}\label{Zmag0}
f\left(Z\right) =
f\left(0\right)\prod_{Z^{\prime}=1}^{Z}\left[\frac{J_{pe}\left(Z^{\prime}
-1\right)+J_{ion}\left(Z^{\prime}-1\right)}{J_{e}\left(Z^{\prime}\right)}\right]
\end{equation}

\noindent for $Z>0$, whereas  for $Z<0$ we have

\begin{equation}\label{Zmin0}
f\left(Z\right) =
f\left(0\right)\prod_{Z^{\prime}=Z}^{-1}\left[\frac{J_{e}\left(Z^{\prime}
+1\right)}{J_{pe}\left(Z^{\prime}
\right)+J_{ion}\left(Z^{\prime}\right)}\right]
\end{equation}

\noindent The constant $f\left(0\right)$ is derived from  the
normalization condition

\begin{equation}\label{normalization}
\sum^{\infty}_{-\infty}f\left(Z\right)=1
\end{equation}

\noindent The expressions for $J_{pe}$, $J_{ion}$ and $J_{e}$ are
taken from \citet{Weingartner01b}. As both neutral and ionized PAHs
are present, we indicate with $\chi_{i} = \chi_{i}\left(a\right)$
the fraction of ionized PAHs of dimension $a$, and with
$Q_{abs}^{IPAH}$ and $Q_{abs}^{NPAH}$ the absorption coefficients of
ionized and neutral PAHs. Including both types of PAHs into eqn.
(\ref{fluxPAH}) we obtain

\begin{eqnarray}
J\left( \lambda \right) &=& \frac{\pi
}{n_{H}hc}\int\nolimits_{\lambda _{min}}^{\lambda _{max}}I \left(
\lambda^{^{\prime }} \right) \lambda^{^{\prime
 }} \int\nolimits_{a_{PAH}^{low}}^{a_{PAH}^{high}} \frac{dn_{PAH}\left(a\right)}{da} \times \nonumber\\
&\times & a^{2} \left[ Q_{abs}^{IPAH} \left( a, \lambda^{^{\prime}}
\right)
S_{ION} \left( \lambda^{^{\prime}} ,\lambda ,a \right) \chi_{i} + \right. \\
& +& \left. Q_{abs}^{NPAH} \left( a, \lambda^{^{\prime}} \right)
S_{NEU} \left( \lambda^{^{\prime}} ,\lambda ,a \right) \left( 1-
\chi_{i} \right) \right] da  d\lambda^{^{\prime}}  \nonumber
\end{eqnarray}

\subsection{The LID model}\label{LID}

\begin{figure}
\psfig{file=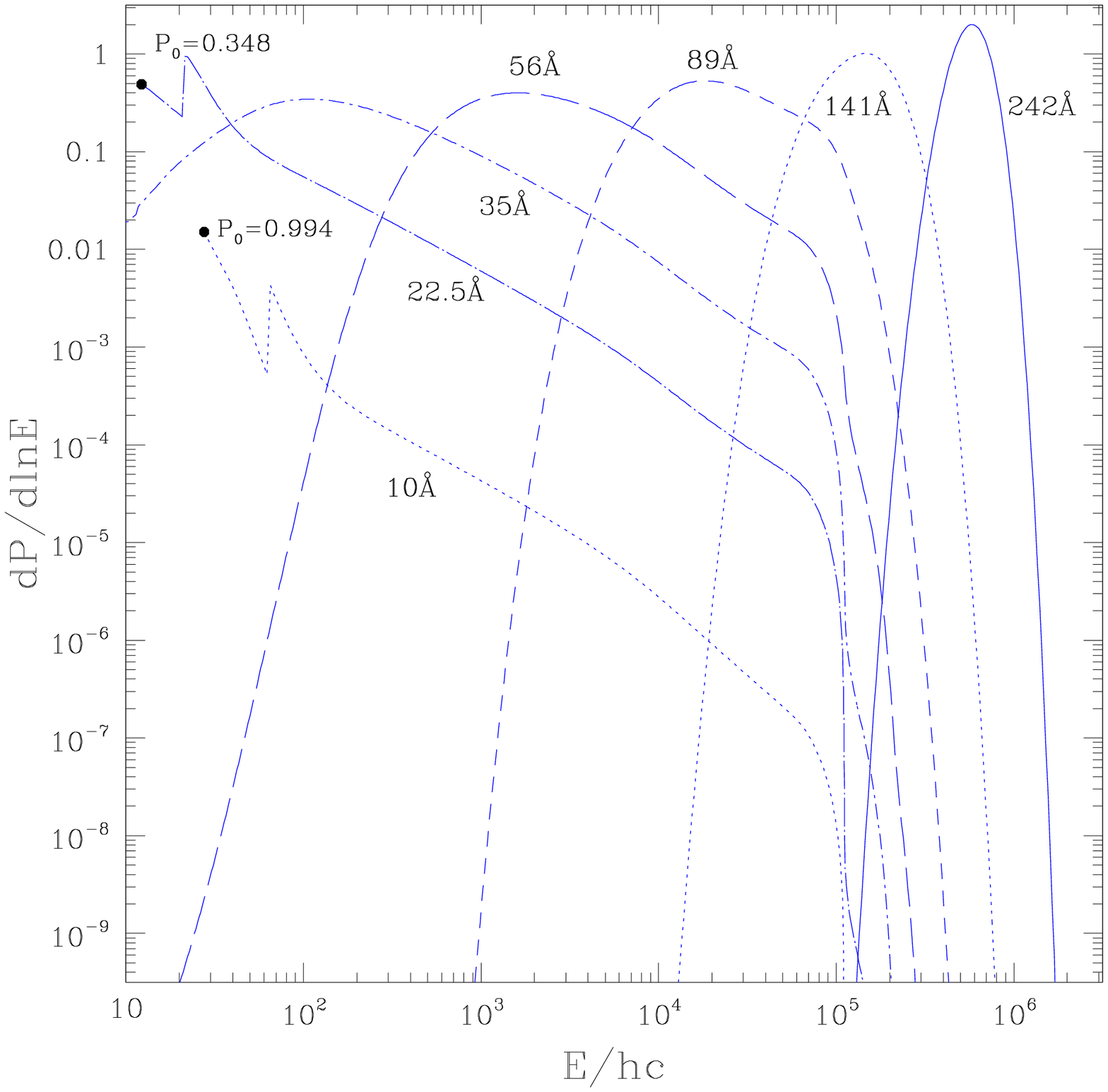,width=8.3truecm} \caption{The energy
distribution functions of silicate grains with dimensions of
$10$\AA, $22.5$\AA, $35$\AA, $56$\AA, $89$\AA, $141$\AA, $242$\AA\
as indicated. All the distributions have been calculated using the
thermal discrete model of \citet{Draine01}. Following \citet{Li01},
we mark with a filled circle the first excited state. $P_{0}$ is the
probability for the grains to be in the ground state that for small
dimensions is high even if the distribution extends to high energy
states.} \label{sildistr}
\end{figure}

We have also implemented in our code the physical and numerical
treatment of the infrared emission of the ISM developed by
\citet{Draine01} and \citet{Li01}, which is currently considered as
the state-of-the art of the subject and has already been used by
\citet{Li02b} to study the IR emission of the SMC. Owing to its
complexity, we will limit ourselves to mention here only a few basic
aspects of the model. For all details the reader should refer to
\citet{Draine01} and \citet{Li01,Li02b}. In this model, made of
three components, i.e. graphite, silicates and PAHs, the energy and
temperature increase of the dust grains caused by the absorption of
energetic photons is calculated by means of the density states of
vibrational energy levels. The task is accomplished in three
possible ways: the exact statistical method, the thermal discrete
approximation and the thermal continuous approach. The latter two
make use of the same energy bins of the exact statistical
approximation, but differ in the way the transitions to higher (by
absorption) and/or lower levels (by emission) are treated. Among the
many improvements upon thermal models made in the past, we call
particular attention on (i) the detailed calculation of vibrational
modes using the harmonic oscillator approximation, (ii) the much
better estimate of the probability of grains to be in the
fundamental state and (iii) the relationship between vibrational
energy and "temperature" of the grains. We have considered here both
types of thermal model, leaving the exact statistical treatment
aside because the enormous computational time required makes it
difficult to implement it in spectrophotometric models of galaxies.
In Fig. \ref{sildistr} we show the energy distribution functions of
the silicates for the thermal discrete model under the effect of the
\citet{Mathis83} radiation field. It is worth noticing how the
probability for the small grains to be in the ground state is high,
even if the distribution extends to high energy states. This
accurate description of the ground state can not be obtained with
the GDP model.

\section{Extinction and emission in the MW, LMC and SMC}\label{ext_emis}

As first step toward including dust emission and extinction in the
SED of young SSPs and galaxies with different shapes, different
histories of star formation and chemical enrichment, different
contents of gas, dust and metals at the present time, we seek to
reproduce the average extinction curves of the MW, LMC and SMC and
to match the emission of their dusty diffuse ISM.
As the MW, LMC and SMC constitute a sequence of decreasing mean
metallicity, they provide the ideal workbench to calibrate the model
for the ISM as a function of the metal content.
Although including the effect of systematic differences in the
metal content is of paramount importance,  this aspect is not
always fully taken into account. For instance \citet{Silva98} adopt
for the diffuse ISM with different chemical compositions a model
originally designed for the MW, and include the effect of different
metallicities by simply scaling the abundance of dust with the
metallicity \citep[see also][for more details]{Devriendt99}.

The procedure we have adopted
iteratively tries to simultaneously satisfy the constraints set by  emission and extinction
 until self-consistency is achieved.
It is also worth mentioning here that  our approach somehow improves upon the analysis by
\citet{Takagi03}, who did not test their dust model on the IR
emission of the SMC. Unfortunately, we cannot test the model
predictions for the emission of PAHs in the LMC because no data are
available \citep[see also][]{Takagi03}. Nevertheless, making use of
COBE and IRAS data, and coupling the fits of the emission and
extinction curves we can put some constraints on the flux intensity
and the relative weight of the dust components.

We  present results obtained with  the various models to  our
disposal, i.e. the  GDP and  LID model for emission and
the WEI and  our MRN-like distribution for the grain sizes. By comparing results from each
possible combination we look for
the  one best suited to model the SED of  SSPs and galaxies.

\begin{figure}
\psfig{file=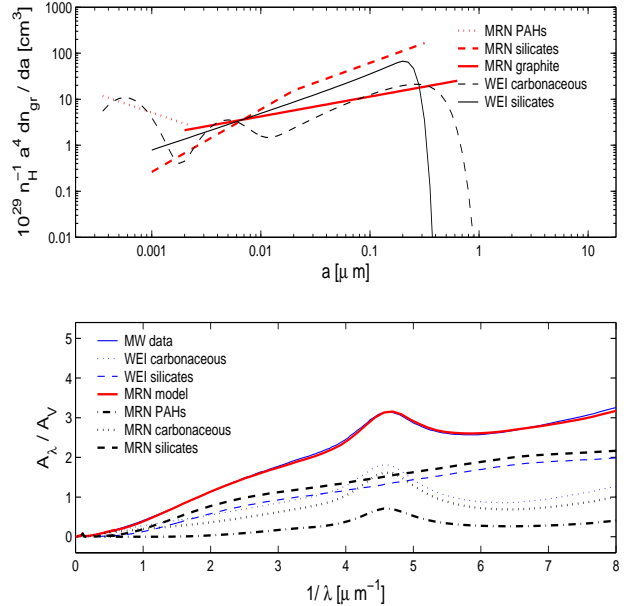,width=8.2truecm,height=8.2truecm}
\caption{ {\bf Top Panel}: The normalized distribution laws (number
density as a function of dimensions) for different types of grain.
For the MRN model we show the graphite grains (thick continuous
line), silicate grains (thick dashed line) and PAHs (thick dotted
line). For the WEI model we display the carbonaceous grains (thin
dashed line) and silicate grains (thin continuous line). {\bf Bottom
Panel}: The extinction curve for the diffuse ISM of the MW obtained
with a mixture made of graphite, silicates and PAHs. In this panel
several cases are shown. First of all the total extinction for our
MRN-like model (thick solid line) and for the WEI model (thin solid
line) together with the observational data (same thin solid line).
All the three lines are practically coincident. In addition to this
for each case we show the contribution from the different types of
grains: the silicates for the MRN and WEI models are indicated with
the thick dashed and the thin dashed lines, respectively; the
contribution of  graphite in the MRN model and of  carbonaceous
grains in the WEI model are displayed by thick dotted and thin
dotted lines, respectively; finally, for MRN model the contribution
of PAHs is shown by the thick dot-dashed line.} \label{estMW}
\end{figure}

\begin{figure}
\psfig{file=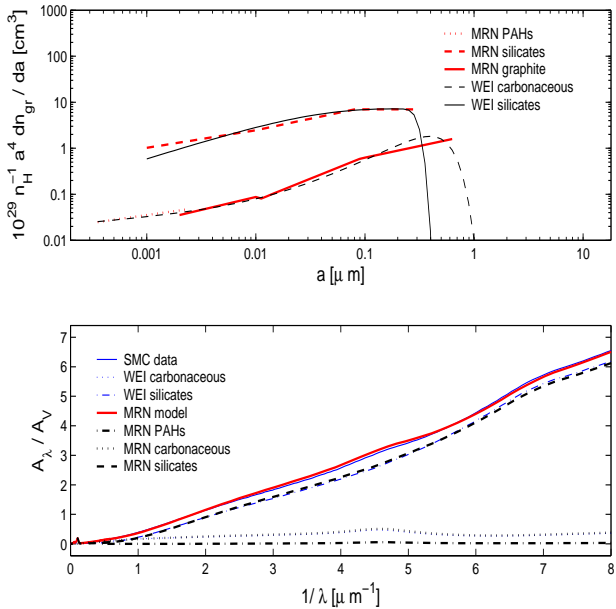,width=8.2truecm,height=8.2truecm}
\caption{The same as in Fig. \ref{estMW} but for the SMC. The data
for the extinction curve of the SMC along the line of sight of the
star AzV398 are taken from \citet{Gordon98}.} \label{estSMC}
\end{figure}

\begin{figure}
\psfig{file=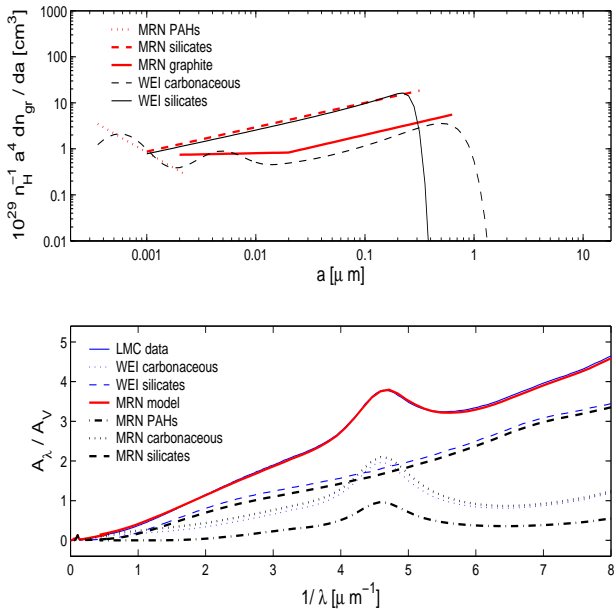,width=8.2truecm,height=8.2truecm}
\caption{The same as in Figs. \ref{estMW} and \ref{estSMC} but for
the LMC. The data for the extinction curve of the LMC are from
\citet{Misselt99}.} \label{estLMC}
\end{figure}

\subsection{The extinction curves of the MW, LMC and SMC}\label{ext_curves}

The best fit of the extinction curves for the three galaxies under
examination is derived from minimizing the $\chi$-square error function
for which we adopt the   \citet{Weingartner01a} definition

\begin{equation}\label{}
\chi^{2}=\sum_{i=1}^{n} \frac{\left(\frac{A_{obs}\left(\lambda_{i}
\right)}{A_{obs}\left(V \right)}-\frac{A_{mod}\left(\lambda_{i}
\right)}{A_{mod}\left(V \right)}\right)^{2}}{\sigma_{i}^{2}}
\end{equation}

\noindent where $A_{obs}\left(\lambda_{i} \right)/A_{obs}\left(V
\right)$ represent the observational data normalized to the V band,
$A_{mod}\left(\lambda_{i} \right)/A_{mod}\left(V \right)$ are the
values of the model, and $\sigma_{i}$ are the weights
\citep{Weingartner01a}. Typically the number of test-points $n$ is
of the order of $100$. We use as optimization technique the
Levenberg-Marquardt method \citep[see][for more details]{Press92}
combined with a sampling algorithm  in bounded hypercubes and some
manual pivoting of the parameters. In any case, it is worth noticing
that the problem is very complicate because the multi-dimensional
solution space is characterized by many local minima to which the
solution may converge missing the true minimum.

In Table (\ref{tableEXT}), we list the parameters we have found from
the best-fit of the extinction curves of the MW, LMC, and SMC,
respectively, using the MRN model. The WEI model is also shown for
comparison.

The MRN-like distribution given by eqn. (\ref{distribnew}) has been
used with $n = 2$ or $n = 3$ for graphite and silicates and $n=2$
for PAHs for the MW and LMC, whereas for the SMC a more complicated
relationship has been adopted. Some parameters are kept
fixed for all the models: i.e. the lower limits for the size distribution
of PAHs and silicates which are set at the lowest values for which
the cross sections are available, $3.5$\AA\ and $10$\AA,
respectively, and the upper limit of the silicates and graphite
distribution fixed at $0.3\,\mu m$ and about $1\,\mu m$. An
important parameter is the upper limit of the PAH distribution,
which  is coincident with the lower limit of the graphite
distribution: the two populations do not overlap and the PAHs are
considered as the small-size  tail of the carbonaceous grains. The
transition size has been fixed at $20$\AA\ thus allowing both
PAHs and small, thermally fluctuating, graphite grains to concur to
shape the SED in the MIR. The exact value of the transition
size has in practice no influence on the extinction curve
thanks to the similar UV-optical properties of PAHs and small
graphite grains. The same considerations do not clearly apply to the
IR emission, where a large population of PAHs will produce stronger
emission in the UIBs. On the contrary, in \citet{Weingartner01a} and
\citet{Li01}, PAHs and graphite grains are included in the family of
the so called "carbonaceous grains", where the smallest grains have
the PAHs optical properties, the biggest grains have the graphite
properties, and for the intermediate dimensions a smooth transition
from PAHs to graphite properties is adopted.

\begin{table*} \label{tableEXT}
\centering \caption{Best fit parameters for the extinction curve of
the diffuse ISM of the MW with $R_{V}=3.1$, for the average
extinction curve of the LMC and for the flat extinction curve of the
SMC. For MW graphite and PAHs we adopt $k=2$, whereas for silicates
we use $k=3$. For LMC silicates and PAHs we adopt $k=2$, whereas for
graphite we use $k=3$. Finally, in the case of SMC, for PAHs we
adopt $k=2$, whereas for graphite we use $k=4$ because we need a
more complex set of power-law distributions.}
\begin{tabular*}{0.88 \textwidth}{lccccccccc}
  \hline
  &    &  MW  &     &     & LMC &    &    &   SMC  &    \\
  \hline
  & Graphite & Silicates & PAHs & Graphite & Silicates& PAHs & Graphite & Silicates & PAHs \\
  \hline
  $A $         & $-25.79$        &   $-25.19$          & $-25.18$          & $-25.96$             &   $-25.35$            & $-25.26$           &  $-26.68$            &   $-25.58$          & $-26.73$           \\
  $\beta    $  & $-3.57$         &   $-3.54$           & $-3.5$            & $-3.5$               &   $-3.47$             & $-3.37$            &  $-3.495$            &   $-3.5$            & $-3.48$            \\
  $\beta_{1}$  &   --            &   $0.9$             & $-1.3$            & $-0.5$               &   --                  & $-2$               &  $0.058$             &   $-0.117$          & $-0.18$           \\
  $\beta_{2}$  &   --            &   $0.4$             &   --              &   --                 &   --                  &   --               &  --                  &   $-0.5$            &   --                   \\
  $\beta_{3}$  &  --             &   --                &   --              &  --                  &   --                  &   --               &  --                  &   --                &   --              \\
  $a_{c_{1}}$\footnotemark[1]  & $2\cdot10^{-7}$ &   $1\cdot10^{-7}$   & $3.5\cdot10^{-8}$ & $2\cdot10^{-7}$      &   $1\cdot10^{-7}$     & $3.5\cdot10^{-8}$  &  $2\cdot10^{-7}$     &   $1\cdot10^{-7}$   & $3.5\cdot10^{-8}$  \\
  $a_{c_{2}}$\footnotemark[1]  & $1\cdot10^{-4}$ &   $2\cdot10^{-6}$   & $2\cdot10^{-7}$   & $2\cdot10^{-6}$      &   $3\cdot10^{-5}$     & $2\cdot10^{-7}$    &  $1\cdot10^{-6}$     &   $1\cdot10^{-6}$   & $2\cdot10^{-7}$   \\
  $a_{c_{3}}$\footnotemark[1]  &   --            &   $3\cdot10^{-5}$   &   --              & $1\cdot10^{-4}$      &   --                  &   --               &  $9\cdot10^{-6}$     &   $8\cdot10^{-6}$   &   --                   \\
  $a_{c_{4}}$\footnotemark[1]  &  --             &   --                &   --              &  --                  &   --                  &   --               &  $1.05\cdot10^{-4}$  &   $3\cdot10^{-5}$   &   --               \\
  $a_{b_{1}}$\footnotemark[1]  &   --            &   $2\cdot10^{-6}$   &   --              & $2\cdot10^{-6}$      &   --                  & $2\cdot10^{-7}$    &   --                 &   $1\cdot10^{-6}$   & $3.5\cdot10^{-8}$   \\
  $a_{b_{2}}$\footnotemark[1]  &   --            &   $2\cdot10^{-6}$   &   --              &   --                 &   --                  &   --               &  $9\cdot10^{-6}$     &   $8\cdot10^{-6}$   &   --                    \\
  $a_{b_{3}}$\footnotemark[1]  &   --            &   --                &   --              &   --                 &   --                  &   --               &   --                 &   --                &   --            \\
  \hline
\end{tabular*}
\begin{minipage}{\textwidth}
\footnotesize$^{1}${All the dimensions are in cm.} \\
\end{minipage}
\end{table*}

The final results for the extinction models,  are shown in Figs.
\ref{estMW}, \ref{estSMC}, and \ref{estLMC} for the MW, SMC and LMC,
respectively. The top panels show the scaled abundances of the
various components in both the MRN- and WEI-models, i.e. PAHs,
graphite and silicates in the former, carbonaceous and silicates in
the latter. The bottom panels show the corresponding extinction
curves, first separately for each component, then the total for both
the MRN- and WEI-models, and finally the observational data.
The MRN- and WEI-models
coincide and both are nearly indistinguishable from the
observational data. Furthermore, it is worth noticing that in order
to obtain good fits for the MW, LMC and SMC the grain size
distributions need to be more complicate than the simple power law
by \citet{Mathis77}. Both in MRN- and WEI- models, the number of
free parameters is about $10$. If a single power-law is adopted as in \citet{Mathis77}
to reduce the number of parameters, we
fail to get a good fit for all the three extinction curves. The
worst situation is with the SMC. The absence of the bump in the
extinction curve makes it difficult to get a good fit without
increasing the number of adjustable parameters. The missing
bump is indeed a strong constraint. Finally, our MRN-model best
solution tends to yield abundances and overall results in close
agreement with those by \citet{Weingartner01a}.

\subsection{Emission by the diffuse ISM of MW, LMC and SMC}\label{emission}

\begin{figure}
\centering
\psfig{file=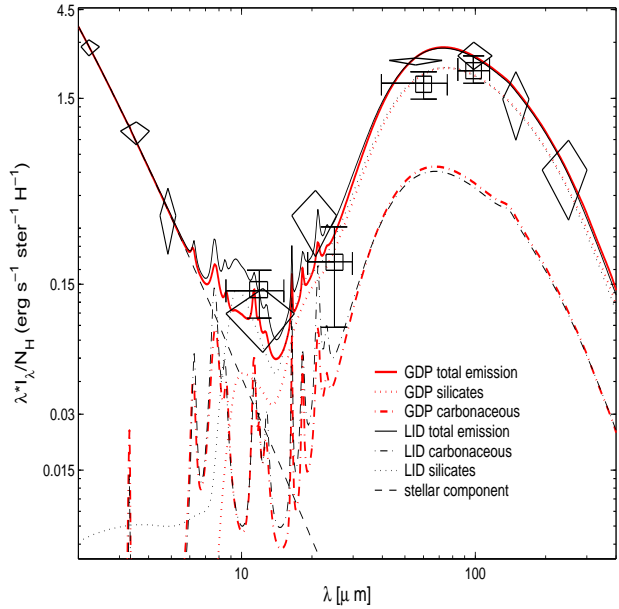,width=8.2truecm,height=8.2truecm}
\caption{Emission of the diffuse ISM predicted by models containing
graphite, silicates and PAHs and comparison with the observational
data for the SMC Bar. The thick lines show the GDP emission model
coupled with our MRN-like grain distribution for SMC. The  thin
lines display the thermal discrete LID emission model of
\citet{Li01} and \citet{Draine01} coupled with the WEI distribution
for SMC \citep{Weingartner01a}. The separate contributions from
different grains are also shown: the solid lines are the total
emission; the dotted lines are the emission of silicates; the
dot-dashed line are the ones of carbonaceous grains; finally the
contribution of the stellar component is indicated by the dashed
line. The incident radiation field is the standard MMP
\citep{Li02b}. The data  from COBE/DIRBE (diamonds) and  IRAS
(squares) have been kindly provided by Li (2004, private
communication). See also \citet{Li02b}.} \label{emissione_SMC}
\end{figure}

Matched the extinction curves, we also need to test the overall
consistency of the model as far as the IR emission by dust and the
comparison with the observational data for the SMC and MW are
concerned. For LMC no data are available and the \citet{Li01} model
is taken as reference model.

The two models of emission (GDP and LID) presented in Sects.
\ref{GDP} and \ref{LID} must be coupled with the two models for the
grain distribution (MRN and WEI) both of which fairly reproduce the
extinction curves as amply discussed in Sect. \ref{ext_curves}.

The combination of WEI distribution with the LID model would be the
ideal case to deal with \citep{Li01,Draine01}. However, we limit
ourselves to consider only the cases GDP+MRN and GDP+WEI, because the
GDP model is much faster than  LID  and therefore more
suited to spectrophotometric synthesis. The LID
model will be, however, considered as the reference case for comparison.
The guide-line here is to cross check results for extinction and
emission and by iterating the procedure to contrive the parameters
at work so that not only unrealistic solutions are ruled out (that
could originate by sole fit of the extinction curve) but also
additional information on the overall problem is acquired.

The preliminary step to undertake is to choose the radiation field
heating up the grains. We adopt the \citet{Mathis83} interstellar
radiation field (MMP) for the solar neighborhood of the MW, both for
the whole MW and the LMC, whereas a slightly different radiation field  is
adopted for the SMC Bar. As this region presents a large spread in
radiation intensities, following \citet{Li02b} and \citet{Dale01},
the intensity of the radiation impinging on the dust grains has been
described with a power-law distribution $dN_{H}/dU$ of MMP radiation
fields \citep{Li02b}.

\begin{figure*}
\centerline{
\psfig{file=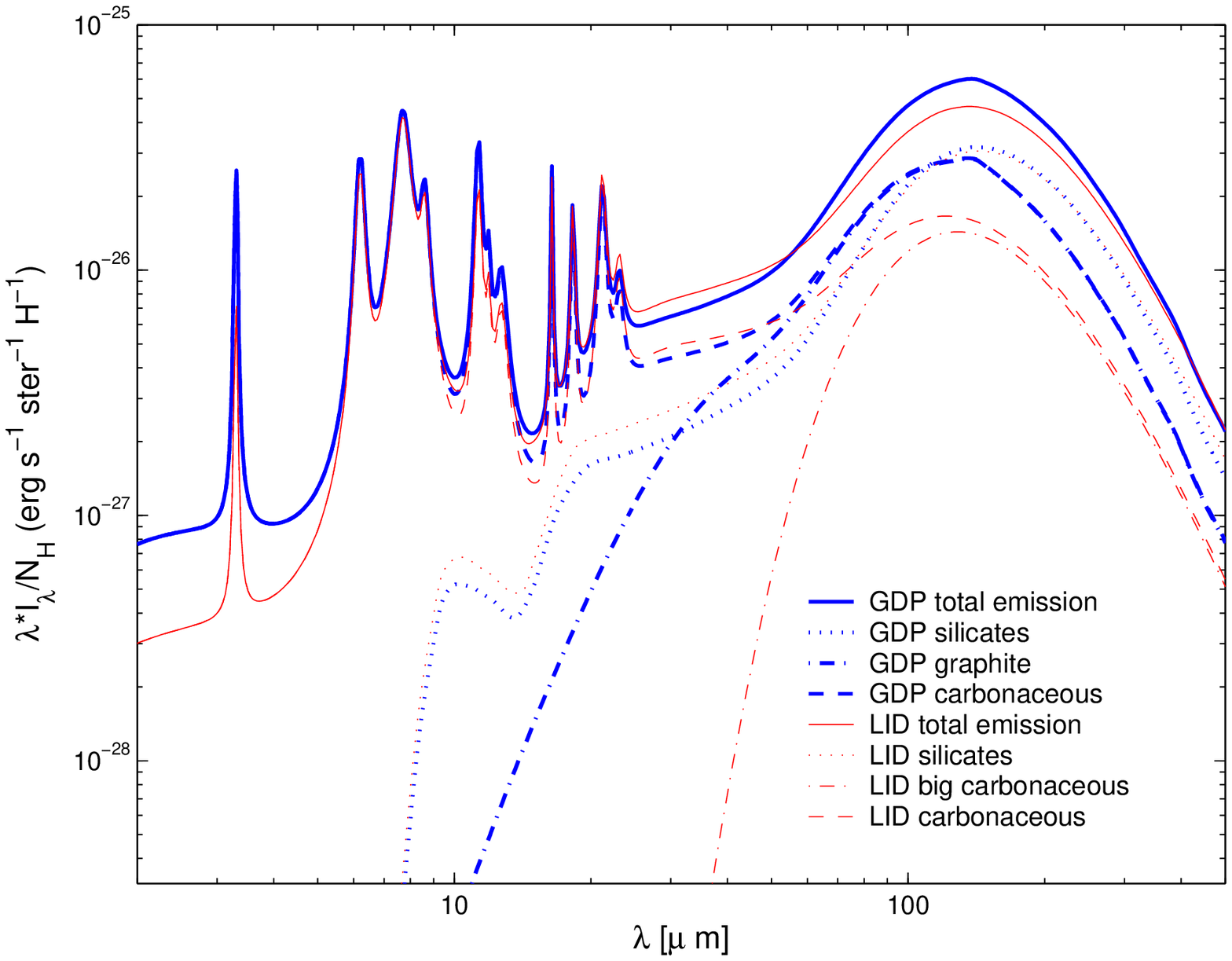,width=8.2truecm,height=8.2truecm}
\psfig{file=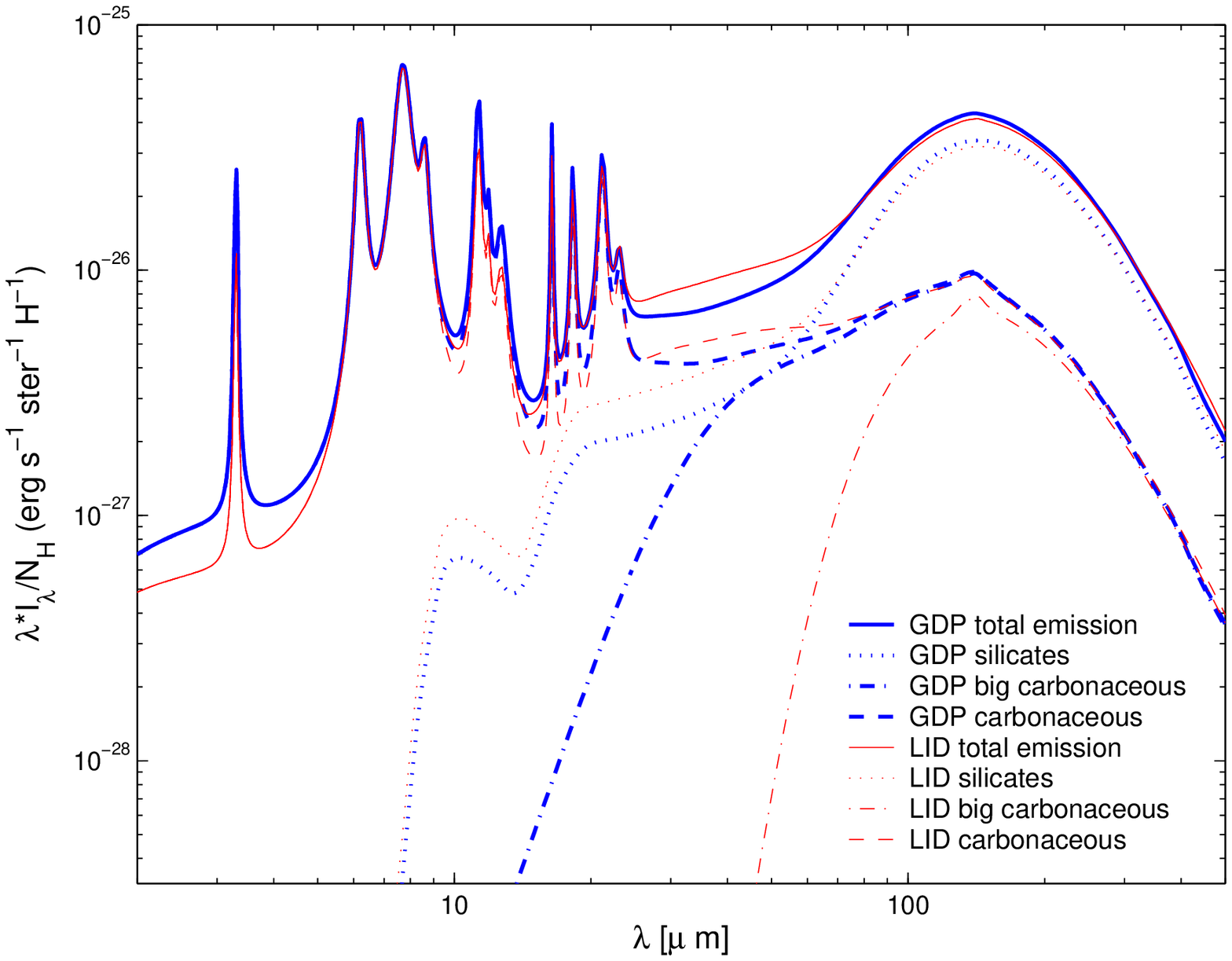,width=8.2truecm,height=8.2truecm}}
\caption{{\bf Left Panel}: Emission of the diffuse ISM predicted by
models containing graphite, silicates and PAHs appropriate to the
case of the LMC. The thick lines correspond to the GDP emission
model with our prescription for MRN-like distribution law suited to
the LMC. The thin lines are the thermal discrete LID emission model
of \citet{Li01} and \citet{Draine01} with the WEI distribution law
\citep{Weingartner01a}. The separate contributions from different
grains are also shown: the solid lines are the total emission; the
dotted lines are the emission of silicates; the dot-dashed lines are
the ones of graphite grains; finally, the short dashed lines are the
total emission of carbonaceous grains, graphite plus PAHs. The
incident radiation field is the standard MMP. {\bf Right Panel}: the
same as in the Left Panel but in all cases the distribution law of
\citet{Weingartner01a} is adopted.} \label{emissione_LMC}
\end{figure*}

\begin{figure*}
\centerline{
\psfig{file=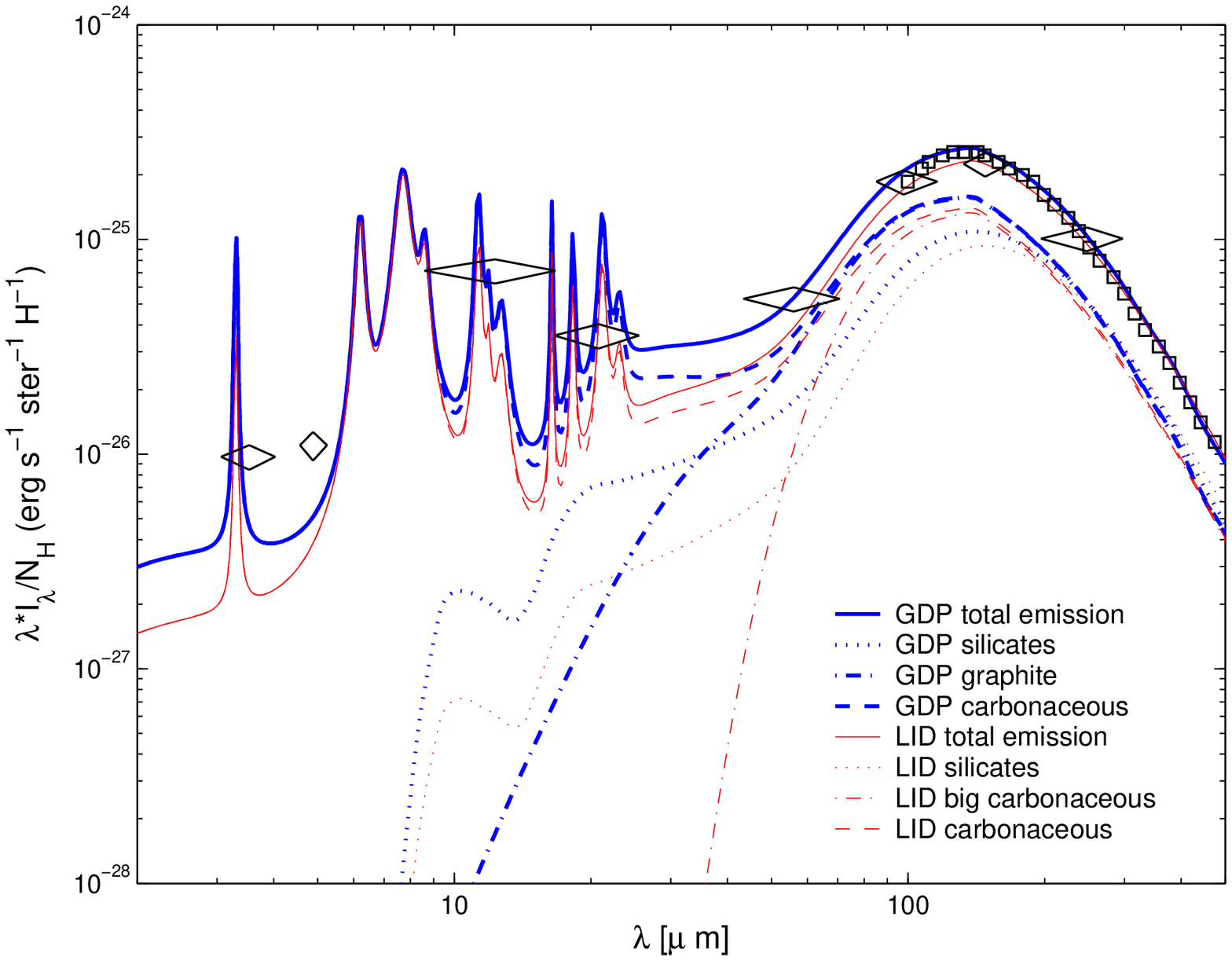,width=8.2truecm,height=8.2truecm}
\psfig{file=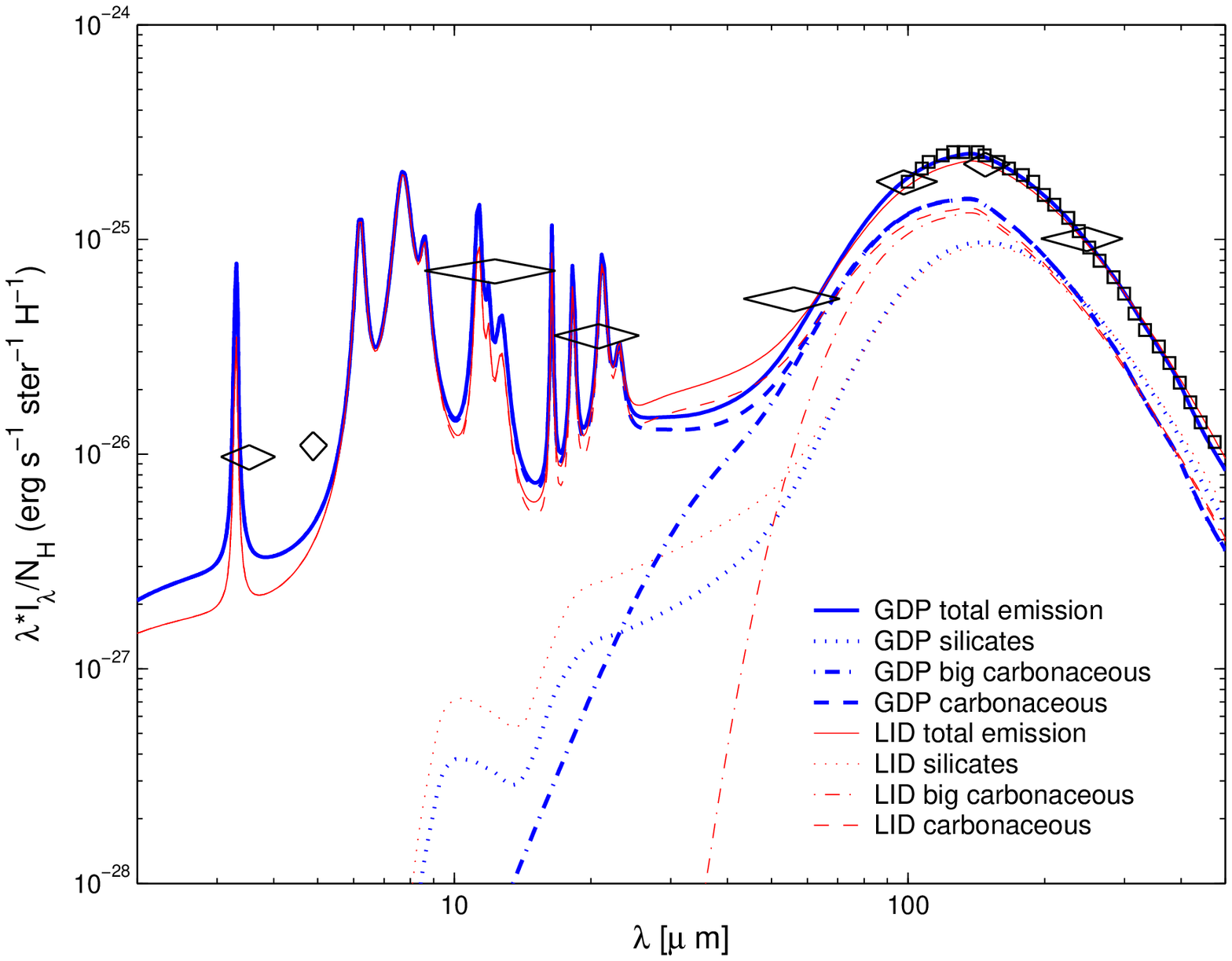,width=8.2truecm,height=8.2truecm}}
\caption{{\bf Left Panel}: Emission of the diffuse ISM in the MW
towards high Galactic latitudes $\left(|b|\geq 25^{\circ}\right)$.
The emission model takes into account graphite, silicates and PAHs.
The thick lines correspond to the GDP emission model with our
prescription for MRN-like distribution law suited to the MW. The
thin lines are the thermal discrete LID emission model of
\citet{Li01} and \citet{Draine01} with the WEI distribution law
\citep{Weingartner01a}. The separate contributions from different
grains are also shown: the solid lines are the total emission; the
dotted lines are the emission of silicates; the dot-dashed line are
the ones of graphite grains; finally, the short dashed lines are the
total emission of carbonaceous grains, graphite plus PAHs. The
incident radiation field is the standard MMP. The observational data
from DIRBE (diamonds) and FIRAS (squares) have been kindly provided
by Li (2004, private communication). See \citet{Li01}. {\bf Right
Panel}: the same as in the Left Panels, but in all cases the same
distribution law of \citet{Weingartner01a} has been adopted.}
\label{emissione_MW}
\end{figure*}

According to \citet{Li02b} the intensity of the IR radiation emitted by the dust
is

\begin{equation}\label{SMCemis}
I_{\lambda}=\frac{1-exp\left[-N_{H}^{tot}\sigma_{abs}\left(\lambda\right)
\right]}{N_{H}^{tot}\sigma_{abs}\left(\lambda\right)}\int^{U_{max}}_{U_{min}}
dU\frac{dN_{H}}{dU}j_{\lambda}\left(U\right)
\end{equation}

\noindent where, $\sigma_{abs}\left(\lambda\right)$ is the total
cross section of absorption, $N_{H}^{tot}$ the total hydrogen column
density and $j_{\lambda}\left(U\right)$ is the emission for the
radiation intensity U, calculated as described in Sect. \ref{GDP}.

In Figs. \ref{emissione_SMC}, \ref{emissione_LMC} and
\ref{emissione_MW} we show the emission expected for the SMC, LMC
and MW, respectively. Results for both GDP+MRN and LID+WEI are
displayed (left panels) together with the observational data,
limited to MW and SMC (the Bar region); data for the LMC are not
available. The comparison highlights the following:

(i) In general the agreement between theory and observational data
is remarkably good. For the SMC, there is a small difference at about
$10 \,\mu m$: this is due to the different distribution of silicates
sizes, which according to \citet{Li02b} may extend down to about
$3.5$\AA, including also nano-particles, for which, however, the
cross sections are not yet available.

(ii) No significant difference between GDP+MRN and LID+WEI can be
noticed.

(iii)  In the case of the SMC the contribution of PAHs to the MIR is very
weak: this follows from the absence of the bump in the average
extinction curve that hints for a very low abundance of carbonaceous
grains, while for the MW and LMC, the emission of PAHs becomes
important.

(iv) The only difference between GDP and LID to note is that the
former overestimates the continuum underneath the $3.3 \,\mu m$
feature. This happens because GDP underestimates the probability of
grains being in the ground state. As a consequence of it, the grains
are generally hotter and thus emit more energy at shorter
wavelengths.

(v) To get a deeper insight on the performance of the GDP model, we
consider the case GDP+WEI and apply it to predict the emission from
MW and LMC. In such a case, both GDP and LID models use the same
grain distribution so that the sole effects of the emission models
can be singled out. The results are shown in the right panels of
Figs. \ref{emissione_LMC} (LMC) and \ref{emissione_MW} (MW). Once
more, not only the GDP model yields results that agree with the
observational data, but also with LID. There are two marginal
differences: one at the $3.3 \,\mu m$ feature and the other at about
$50 \mu m$ where the flux is slightly underestimated. The reason is
always the ground state probability: in GDP grains tend to be hotter
and to emit more at short IR wavelengths thus slightly shifting the
emission from $50 \mu m$ to shorter wavelengths. Finally, using in
both GDP and LID models the WEI distribution, the small differences
noticed for LMC at $100 \mu m$ also disappear.

Basing on this combined analysis of emission and extinction, we
decided to adopt GDP+WEI to describe the dusty ISM and to model the
SED of young dusty SSPs and of galaxy spectrophotometric models that
will be presented in the companion paper by \citet{Piovan05}.

There is a final consideration about our choice for the size
distribution law. Even if both WEI and MRN  lead to good fits, it is
only the emission model that determines  the  correct solution. It
is worth recalling that it is not possible to constrain the $C$
abundance only by fitting the extinction curve   which simply
provides an upper limit \citep[see][for more
details]{Weingartner01a}. With the MRN model it is much more
difficult to obtain good fits of the extinction curves at varying
the $C$ abundance and to couple extinction and emission. In
contrast, with the WEI model, which considers $b_{C}$ as a free
parameter, the goal can be easily achieved. Therefore, the WEI
distribution law is perhaps best suited to model the SEDs of young
dusty SSPs and galaxies.

\section{Population synthesis for SSPs}\label{basic_popsyn}

The monochromatic flux of a SSP of age $t$ and metallicity $Z$ at
the wavelength $\lambda$ is defined as

\begin{equation}
ssp_{\lambda }\left( t,Z \right) =\int\nolimits_{M_{L}}^{M_{U}(t)}
f_{\lambda }\left( M,t,Z \right) \Phi (M) dM \label{def_flux_ssp}
\end{equation}

\noindent where $f_{\lambda }\left( M,t,Z\right) $ is the
monochromatic flux emitted by a star of mass $M$, age $t$, and
metallicity $Z $; $\Phi\left(M\right)$ is the IMF; $M_{L}$ is the
mass of the lowest mass star in the SSP, whereas $M_U(t)$ is mass of
the highest mass star still alive in the SSP of age $t$. For the IMF
we adopt the \citet{Salpeter55} law expressed as $dN/dM = \Phi(M) =
\mathcal{A} M^{-x}$ where $x$=2.35 and $\mathcal{A}$ is a
normalization constant to be fixed by a suitable condition (SSPs for
other choices of the IMF can be easily calculated).

In our study we adopt the isochrones by \citet{Tantalo98a}
(anticipated in the data base for galaxy evolution models by
\citet{Leitherer96}). The underlying stellar models are those of the
Padova Library \citep[see][for more details]{Bertelli94}. The
initial masses of the stellar models go from 0.15 to 120
$M_{\odot}$. The following initial chemical compositions have been
considered: [Y=0.230, Z=0.0004], [Y=0.240, Z=0.004], [Y=0.250,
Z=0.008], [Y=0.280, Z=0.02], [Y=0.352, Z=0.05], and [Y=0.475,
Z=0.1], where $Y$ and $Z$ are the helium and metal content (by
mass).

The stellar spectra in usage here are taken  from the library of
\citet{Lejeune98}, which stands on the Kurucz (1995) release of
theoretical spectra, however with several important implementations.
For $T_{eff}<3500$ K the spectra of dwarf stars by \citet{Allard95}
are included and for giant stars the spectra by \citet{Fluks94} and
\citet{Bessell89,Bessell91} are considered. Following
\citet{Bressan94}, for $T_{eff}> 50000$ K, the library has been
extended using black body spectra.

The SSP spectra have been calculated for the same ranges of ages and
metallicities of the isochrones.

\section{Spatial distribution of young stars, gas and dust}\label{sp_dis_sgd}

The first step to undertake is to specify the relative distribution
of young stars and dust. From the observational maps of MCs it is
soon evident that the situation is very complicate: there are many
point sources of radiation (the stars) enshrouded by dust, that are
randomly distributed across clouds of irregular shape. Inside a
cloud the temperature varies locally depending not only on the
distance $r$ from the cloud center but also on the distance from the
nearest star. In other words a dust grain not particularly close to
a star will have a temperature $T\left( r\right)$ determined by the
local, average interstellar radiation field (ISRF), whereas a dust
grain at the same distance $r$ from the center, but close to a hot
star, will experience a hotter temperature \citep{Krugel94}. All
this implies that the spherical symmetry is broken, thus drastically
increasing the complexity of the whole problem. The assumption of
spherical symmetry is the only way to handle this kind of problem at
a reasonable cost. However, even with spherical symmetry,  the
spatial distribution of stars, gas and dust can be realized in many
different ways. Finally, there are different techniques to deal with
the radiative transfer across the dense medium of the MCs.

\citet{Krugel94} analyzed in a great detail three possible
spherically symmetric configurations: first the case in which hot
stars enshrouded by dust (the so-called "hot-spots") are randomly
distributed \citep{Krugel78}; secondly the case of a central point
source (the "point-source" model) \citep[see][]{Rowan80,Rowan89} and
finally the case with an extended central source
\citep{Siebenmorgen91}.

The hot-spots description (or anyone of the same type) would be
closer to reality and the one providing more physically sounded
results. The back side of the coin is that compared to the central
source case, the problem gets soon complicate
from the point of view of numerical computations. For this reason we
take an hybrid case in between the hot-spots and central
point-source descriptions. We simulate a young stellar population
embedded in a spherical cloud of gas and dust by assuming that
stars, gas and dust follow three different King's laws each of which
characterized by its own parameters:

\begin{equation}
\rho_{i}\left(r\right)=\rho_{0,i}\left[ 1+\left(
\frac{r}{r_{c,i}}\right) ^{2}\right]^{-\gamma_{i}}
\label{rhostar_ell}
\end{equation}

\noindent where the index $i$ can be $s$ (stars), $d$ (dust) and $g$
(gas), $r_{c,s}, r_{c,g}$ and $r_{c,d}$ are the core radii, whereas
$\gamma_{g}, \gamma_{d}$ and $ \gamma_{s}$ are the exponents of the
distributions of gas, dust and stars. As dust and gas are mixed
together, they are described by the same parameters,
$r_{c,d}=r_{c,g}$. The exponents are simply chosen to be
$\gamma_{g}=\gamma_{d}=\gamma_{s}=1.5$. As in \citet{Takagi03}, we
introduce the parameter $\eta=r_{c,d}/r_{c,s}$, equal to the ratio
between the two scales. Following \citet{Combes95} we adopt the
relation $\log\left(r_{t,s}/r_{c,s}\right)=2.2$ between the tidal
radius $r_{t,s}$ and the scale radius of stars $r_{c,s}$ and simply
assume $r_{t,s}=r_{t,g}=r_{t,d}=r_{t}$. Denoting with $R_{MC}$ the
radius of a generic MC, and imposing $r_{t}=R_{MC}$, there will be
no dust and  gas for $r > R_{MC}$, i.e. for $r
> r_{t}$ \citep{Takagi03}.

To fully understand the differences between a central point-source
and a spatially distributed source it may be worth of interest to
compare SSPs calculated with the two schemes. The SSPs for the
spatially distributed source are the ones presented here. They are
calculated using the ray tracing technique to be explained in Sect.
\ref{ray_tracing}. For SSPs based on the central point-source
description we adopt the code DUSTY\footnote{The code is kindly made
available by the authors at \textit{http://www.pa.uky.edu/$ \sim
$moshe/dusty}.}  to solve the radiative transfer equation
\citep{Ivezic97}. Even if DUSTY is not able to handle thermally
fluctuating grains and PAH emission, it is fully adequate to show
the difference arising from  different geometries.

In DUSTY only two quantities must be specified  to obtain a complete
solution: the total optical depth $\tau_{\lambda}$ at a suitable
reference wavelength, and the condensation temperature of the dust
$\left(T_{s}\right)$ at the inner edge of the shell. All other
parameters are defined as dimension-less and/or normalized
quantities. We have fixed $T_{s}$ to a suitable value, following the
suggestion by \citet{Silva98} who made use of central point source
approach to model MCs.

\begin{figure}
\psfig{file=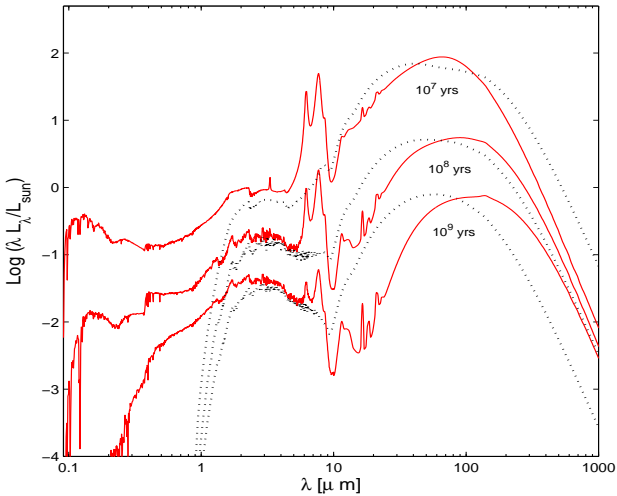,width=8.3truecm,height=6.6truecm}
\caption{Spectra of dusty SSPs of metallicity $Z=0.02$ calculated
with DUSTY (central point source - dotted lines) and with the ray
tracing method (distributed source - continuous lines). They are
represented at three different ages, from top to bottom: $10$ Myr,
$100$ Myr and $1$ Gyr.} \label{RAYDUSTY}
\end{figure}

In Fig. \ref{RAYDUSTY} we compare young dusty SSPs calculated both
with DUSTY (central point-source) and the ray tracing technique
(distributed source). Both SSPs have the solar metallicity, the same
optical depth and matter profile of the cloud, and the same dust
properties. The geometry is the origin of the lack of stellar
emission for the dusty SSPs calculated with the central point-source
model. All the stars are indeed embedded in a medium of high optical
depth so that no stellar light can escape from it. Even if in some
cases of highly embedded sources this could be a good approximation,
in general it is not realistic, because we can easily conceive that
in a real environment with ongoing star formation, young stars can
be formed everywhere even near the edges of the MCs. Such stars are
therefore less obscured by dust. The ray-tracing method with the
stars distributed into the cloud can easily handle this situation
thus allowing a fraction of the stellar light to be always
detectable. This is a net advantage with respect to the central
point-source approximation. As a matter of facts, there is
observational evidence for contributions to the UV-NIR flux from
bursts of young stars  as discussed by \citet[see the discussion
by][]{Takagi03}. The central point-source model is not able to
simulate properly this contribution, which is instead nicely
described by  young stars being distributed all across the cloud.
Anyway, as the point-source model can reasonably simulate highly
obscured and concentrated populations of stars, we plan to introduce
also this case in our library of SEDs.

\section{The radiative transfer problem}\label{rad_transport}

\subsection{The ray tracing method}\label{ray_tracing}

\begin{figure}
\psfig{file=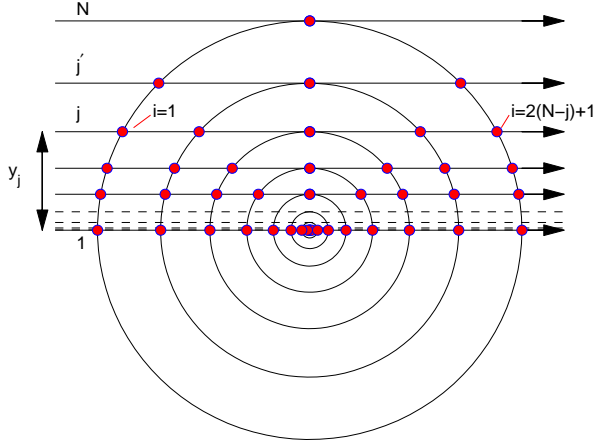,width=8truecm} \caption{Two dimensional
section of a dusty MC. A set of $N$ rays is traced each one tangent
to one of N concentric spherical surfaces of a radial grid with
$N=10$ and thicker spacing toward the center of the cloud. The
specific intensity is calculated for each ray at each point when the
ray penetrates one of the surfaces. The $j$-th ray cross $N-j$
surfaces in $2\left(N-j\right)+1$ points. $y_{j}$ is the impact
parameter of the $j$-th ray.} \label{Circle}
\end{figure}

To solve the radiative transfer equation we use the robust and
simple technique otherwise known as the "ray tracing" method
\citep{Band85,Takagi03}. The method solves the radiative transfer
equation along a set of rays traced throughout the inhomogeneous
spherically symmetric source. Thanks to the spherical symmetry of
the problem, it is possible to calculate the specific intensity of
the radiation field at a given distance from the center of the MC by
suitably averaging the intensities of all the rays passing through
that point \citep{Band85}. The geometry of the problem is best
illustrated in Fig. \ref{Circle} which  shows a section of a
spherically symmetric  MC. We consider a discrete set of $N$
concentric spheres (circles on the projection) and an equal number
of rays indicated by the parameter $j$ running from $j$=1 to $j$=N.
The specific intensity is calculated at all intersections of the
generic ray $j$ with the generic circle (sphere). The number of
intersections $i$ increases from 1 for $j$=$N$ to $2(N-1)+1$
according to the rule $i$=$2(N-j)+1$. The equation of radiative
transfer for specific intensity $I_{\lambda}\left(y,x\right)$ along
a ray can be written as

\begin{equation}\label{radiative_transfer}
\frac{dI_{\lambda}\left(y,x\right)}{dx}=-n_{H}\left(r\right)\sigma_{ext}\left(\lambda\right)
\left[I_{\lambda}\left(y,x\right)-S_{\lambda}\left(r\right)\right]
\end{equation}

\noindent where $y$ is the impact parameter, i.e. the distance of
the generic ray $j$  from the ray passing through the center, and
$x$ is the coordinate along the ray;  $n_{H}\left(r\right)$ is the
number of hydrogen atoms per cubic centimeter;
$\sigma_{ext}\left(\lambda\right)$ is the cross section of
extinction for hydrogen atoms defined in eqn. (\ref{sigabs});
finally, $S_{\lambda}\left(r\right)$ is the source function at a
given radius. We include the metallicity dependence for the
composition of the MCs (on the notion that a generation of newly
born stars shares the same metallicity of the surrounding MCs) as
follows: for metallicities $Z \geq 0.015$ we use the extinction
curve of the MW for dense regions, characterized by a reddening
parameter as high as is $R_{V}=5.5$; for metallicities in the
interval $0.005 \leq Z \leq 0.015$ we adopt the extinction curve of
the LMC; finally, for metallicities lower than  $Z \leq 0.005$ we
use the typical extinction curve of the SMC.

In order to solve eqn. (\ref{radiative_transfer}),  following
\citet{Rowan80} we split  the specific intensity in three
components:
$I_{\lambda}=I_{\lambda}^{\left(1\right)}+I_{\lambda}^{\left(2\right)}+I_{\lambda}^{\left(3\right)}$,
where $I_{\lambda}^{\left(1\right)}$ is the  light from the stellar
source, $I_{\lambda}^{\left(2\right)}$ is the light scattered  by
the grains, and, $I_{\lambda}^{\left(3\right)}$ is the radiation
coming from the thermal emission of the grains. It is possible to
calculate the contribution from each component, if the optical
properties of the dust grains do not depend on their temperature.
This is indeed the case of the optical coefficients taken from
\citet{Draine84}, \citet{Laor93} and \citet{Li01}. As we are in
spherical geometry only a radial grid needs to be specified. For
each impact parameter $y_{j}=r_{j}$, $I_{\lambda}^{\left(1\right)}$
is calculated and stored at each intersection of the $j$-th ray with
the spheres (circles) (as   shown in Fig. \ref{Circle}), where
the stellar source function
$S_{\lambda}^{\left(1\right)}\left(r\right)$ is taken from
\citet{Takagi03}.

\begin{figure}
\psfig{file=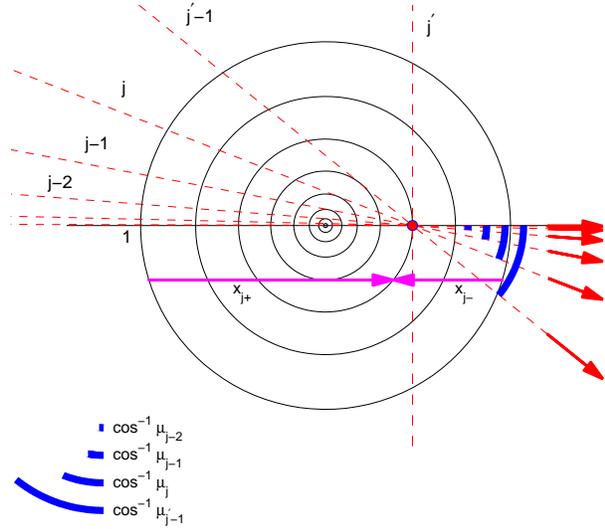,width=8truecm} \caption{Two dimensional
section of a dusty MC. A set of $N$ spherical surfaces of a radial
grid with $N=10$ is shown. The average intensity is calculated for
the point $j^{\prime}$ by using all the rays from $1$ to
$j^{\prime}-1$, each rays weighted for the cosine of the angle with
the radial vector. For each ray we have two contributes, one for
each direction, $x_{j_{-}}$ and $x_{j_{+}}$.} \label{Circle2}
\end{figure}

Once the matrix
$I_{\lambda}^{\left(1\right)}\left(y_{j},x_{i+1}\right)$ is
calculated, it is possible to obtain the mean intensity
$J_{\lambda}^{\left(1\right)}\left(r_{j}\right)$ at a given radius
performing an integration over the cosine grid $\mu_{j}$ as
displayed in Fig. \ref{Circle2}. We have now to calculate the
contribution of $I_{\lambda}^{\left(2\right)}$ and
$I_{\lambda}^{\left(3\right)}$ to $J_{\lambda}$. The source function
of the $n$-times scattered light is

\begin{equation}\label{scattering}
S_{\lambda,n}^{\left(2\right)}\left(r,\overrightarrow{n}\right)=\frac{\omega_{\lambda}}{4\pi}
\int
\phi\left(\overrightarrow{m},\overrightarrow{n}\right)I_{\lambda,n-1}^
{\left(2\right)}\left(r,\overrightarrow{m}\right)d\Omega
\end{equation}

\noindent where $I_{\lambda,n-1}^{\left(2\right)}$ is the associated
specific intensity. The function
$\phi\left(\overrightarrow{m},\overrightarrow{n}\right)$ is the
phase function, which yields the distribution of photons scattered
in different directions. For the sake of simplicity we assume
isotropic scattering so that
$\phi\left(\overrightarrow{m},\overrightarrow{n}\right)=1$. The
total intensity of scattered light is obtained by adding all
scattering terms
$I_{\lambda}^{\left(2\right)}=\sum_{n}I_{\lambda,n}^{\left(2\right)}$.
We start from  $I_{\lambda}^{\left(1\right)}$, which can  be
considered as the $0$-times scattered light, to obtain the source
function of $1$-times scattered light \citep{Takagi03} and the
average specific intensity for each spherical concentric shell. We
derive $S_{\lambda,n}^{\left(2\right)}\left(r\right)$ and repeat
these steps, calculating the scattering terms of higher order. By
taking a sufficiently large number of steps, six  are sufficient \citep{Takagi03}, the energy
conservation is secured.

Finally, we need  the source function for the infrared emission of
dust. We take also dust self-absorption into account: for high
optical depths such as the ones of MCs, the infrared photons can be
absorbed and self-absorption has to be considered. An iterative
procedure is required until the energy conservation is reached:
using the mean intensities of direct and scattered light,
$J_{\lambda}^{\left(1\right)}\left(r\right)$ and
$J_{\lambda}^{\left(2\right)}\left(r\right)$, the dust source
function for the $1$-th iteration
$S_{\lambda,1}^{\left(3\right)}\left(r\right)$ together with the
corresponding average intensity
$J_{\lambda,1}^{\left(3\right)}\left(r\right)$ are obtained.
Starting from $J_{\lambda,1}^{\left(3\right)}\left(r\right)$, we
iterate the procedure to get the thermal emission of dust
$J_{\lambda,n}^{\left(3\right)}\left(r\right)$ by means of the
 the local radiation field
$J_{\lambda}^{\left(2\right)}\left(r\right)+J_{\lambda}^{\left(1\right)}\left(r\right)+
J_{\lambda,n-1}^{\left(3\right)}\left(r\right)$. The dust emission
is calculated as described in Sect. \ref{GDP} with the GDP model.
Therefore, the source function of dust is given by

\begin{eqnarray}
S_{\lambda,1}^{\left(3\right)}\left(r\right)=\frac{1}{\sigma_{ext}\left(\lambda
\right)}\left(j_{\lambda}^{small}+j_{\lambda}^{big}+j_{\lambda}^{PAH}\right)
\end{eqnarray}

\noindent where $j_{\lambda}^{small}$, $j_{\lambda}^{big}$ and
$j_{\lambda}^{PAH}$ are the contributions to the emission by
small grains, big grains and PAHs, respectively.

\subsection{Optical depth of the cloud}\label{opt_depth}

Key parameter of the radiative transfer is the optical depth
$\tau_{\lambda}$ defined as

\begin{equation}
\tau_{\lambda }=\int\nolimits_{S}k_{\lambda}\rho ds
\label{taulambda}
\end{equation}

\noindent where $k_{\lambda}$ is the extinction coefficient, $\rho$
is the density of the matter and $S$ is the thickness of the
cylinder of matter along which we integrate. Eqn. (\ref{taulambda})
can be recast as follows

\begin{equation} \label{tau_1}
\tau_{\lambda}=\int_{0}^{R_{MC}}n_{H}\left(r\right)\sigma_{\lambda}\left(r\right)dr
\end{equation}

\noindent where $n_{H}\left(r\right)$ is the number density of $H$
atoms and $\sigma_{\lambda}\left(r\right)$ is the extinction cross
section (scattering plus absorption).

In order to calculate the thickness of the shell, we need an
estimate of the mass $M_{MC}$ and the radius $R_{MC}$ of the cloud.
\citet{Silva98} adopt a "typical" MC of given mass and radius.
However, there is plenty of evidence that MCs span large ranges of
masses and radii in which the low mass and small radius ones are by
far more numerous. Therefore, in a realistic picture of the effects
of dust clouds, the mass and radius distributions of these latter
should be taken into account thus requiring that large ranges of
optical depths are considered. We start calculating the optical
depth for a cloud with a generic mass $M_{MC}$ and radius $R_{MC}$.
From the identity
$\rho_{g}\left(r\right)=\rho_{H}\left(r\right)+\rho_{He}\left(r\right)+\rho_{Z}\left(r\right)$
with $\rho_{Z}\left(r\right)$, $\rho_{H}\left(r\right)$,
$\rho_{He}\left(r\right)$ densities of metals, hydrogen, and helium,
respectively, after  putting $\rho_{H}\left(r\right)$ into evidence
and neglecting $\rho_{Z}\left(r\right)/\rho_{H}\left(r\right)$ we
get:

\begin{equation}
\rho_{H}\left(r\right) \sim \frac{\rho_{g}\left(r\right)}{\left(
1+\frac{\displaystyle\rho_{He}\left(r\right)}{\displaystyle\rho_{H}\left(r\right)}\right)}
\sim \rho_{0,g}\frac{\left[ 1+\left( \frac{\displaystyle
r}{\displaystyle r_{c,g}}\right) ^{2}\right]^{-\gamma_{g}}}{\left(
1+
\frac{\displaystyle\rho_{He}\left(r\right)}{\displaystyle\rho_{H}\left(r\right)}\right)}
 \label{hdensita}
\end{equation}

\noindent We can now reasonably assume that the hydrogen and helium
distributions have the same scale radius, so that
$\rho_{He}\left(r\right)/\rho_{H}\left(r\right) =
\rho_{He,0}/\rho_{H,0}$ does not  depend on  the coordinate $r$.
Dividing eqn. (\ref{hdensita}) by the mass of the $H$ atom $m_{H}$ and
supposing  that $\sigma_{\lambda}\left(r\right)$ does not depend on
the coordinate $r$ along the cylinder of matter we derive

\begin{equation}
\tau_{\lambda}=\frac{\sigma_{\lambda}\rho_{0,g}}{\left( 1+
\frac{\displaystyle\rho_{He,0}}{\displaystyle\rho_{H,0}}\right)m_{H}}\int_{0}^{R_{MC}}\left[
1+\left( \frac{r}{r_{c,g}}\right) ^{2}\right]^{-\gamma_{g}}dr
\end{equation}

\noindent Setting $z=r/r_{c,g}$ and introducing the parameter
$\eta=r_{c,d}/r_{c,s}=r_{c,g}/r_{c,s}$  -- with the ratio of the
cloud radius $R_{MC}$ to the scale radius of the gas component given
by $R_{MC}/r_{c,g}=10^{2.2}\cdot(r_{c,s}/r_{c,g})=10^{2.2}/\eta$ --
and $\gamma_{g} = 1.5$ we obtain

\begin{equation}
\tau_{\lambda}=\frac{\sigma_{\lambda}\rho_{0,g}R_{MC}}{\left( 1+
\frac{\displaystyle\rho_{He,0}}{\displaystyle\rho_{H,0}}\right)m_{H}}
\times \frac{\eta}{10^{2.2}}\int_{0}^{\frac{10^{2.2}}{\eta}}\left(
1+z^{2}\right)^{-1.5}dz \label{tau_4}
\end{equation}

\noindent The normalization constant $\rho_{0,g}$ is derived by
integrating the density law for a MC given by eqn.
(\ref{rhostar_ell}) over its volume and using the identity

\begin{equation}
M_{MC}= 4\pi \, \rho_{0,g} \,(r_{c,g})^3 \int_0^{R_{MC}/r_{c,g}}
{\frac{x^2}{(1+x^2)^{-1.5}}} dx \label{norma_mc}
\end{equation}

\noindent with $x=r/r_{c,g}$.

The above expression for the optical depth depends only on suitable
constants and the dimension-less integral which is a function of
$\eta$. Denoting with

\begin{equation}\label{phi_eta}
\Phi\left(\eta\right)=
\frac{\eta}{10^{2.2}}\int_{0}^{\frac{10^{2.2}}{\eta}}\left(
1+z^{2}\right)^{-1.5}dz
\end{equation}

\noindent and $K=1/\left[\left(1+
\displaystyle\rho_{He,0}/\displaystyle\rho_{H,0}\right)\,
m_{H}\right]$, we finally get

\begin{equation}\label{tau_final}
\tau_{\lambda}=K\sigma_{\lambda}\rho_{0,g}R_{MC} \,
\Phi\left(\eta\right)
\end{equation}

\noindent The parameter $\eta$ describes the dust distribution: for
$\eta=1000$ we have a nearly uniform distribution of dust in which
stars are embedded, whereas for $\eta=1$ stars and dust in the MCs
have the same distribution. Typical values of
$\Phi\left(\eta\right)$ are
$\Phi\left(\eta\right)=0.0063,0.063,0.534,0.988$ as $\eta$ goes from
1 to 10, 100, 1000, respectively. Following \citet{Takagi03} who
reproduced the central star forming regions of M82 and Arp220 using
a nearly uniform distribution of dust ($\eta=1000$), we adopt the
same value for this parameter. Looking at the optical depth of the
cloud we have just defined, there are other considerations to be
made about the geometry of the system. Both the mass $M_{MC}$, via
the central density $\rho_{0,g}$, and the radius $R_{MC}$ of the
cloud are present in  eqn. (\ref{tau_final}). The first problem to
deal with is that masses and radii of real MCs span large  ranges of
values.  This problem will be addressed in detail in Sect.
\ref{MCs_distribution}. Second,  in all the relationships we have
been using in so far to define the final expression for the optical
depth,  the total mass of the stars embedded into the MC does not
play any role. This is obvious because the optical depth depends
only on the dust-fog in which the stars-candles are embedded, and
not on the candles themselves. So, which star mass $M_{\star}$ has
to be considered for a cloud with mass $M_{MC}$ and radius $R_{MC}$?
There is no easy answer to this problem. We start considering  that
our MCs spectra will be eventually used to build integrated spectra
of galaxies. Therefore, we need normalized MCs spectra to be
convolved with the SFH of the galaxy. The simplest way we can
achieve the goal is to consider as candle a SSP of $M_{\star}=1
M_{\odot}$. We keep fixed the optical depth of the cloud calculated
with eqn. (\ref{tau_final}) using the real dimensions of the cloud
and then we re-scale the radius of the cloud according to the
relation \citep{Takagi03}

\begin{equation}\label{scaled_radius}
R_{MC}^{\prime}=10R_{s}\left(\frac{M_{\star}}{10^{11}M_{\odot}}\right)^{0.5}
\end{equation}

\noindent where $R_{MC}^{\prime}$ (in kpc) is the normalized radius
corresponding to the adopted mass $M_{\star}$ (for an arbitrary star
mass $M_{\star}$, the radius $R_{MC}$ has to be suitably scaled). As
noticed by  \citet{Takagi03}, when the optical depth and the scaling
parameter $R_{s}$ are kept fixed, the relative shape of the SED
remains unchanged for different masses $M_{\star}$. The only effect
of varying $M_{\star}$ is to produce a higher energy output, but the
SED remains the same.

\subsection{The mass spectrum of MCs}\label{MCs_distribution}

As already mentioned, there is observational evidence that the MCs
span large ranges of masses and radii. Let us assume now that masses
and radii of molecular clouds obey the distribution laws
$dN/dM_{MC}=A_{M}M_{MC}^{-\alpha}$ and
$dN/dR_{MC}=A_{R}R_{MC}^{-\beta}$ where the normalization constants
$A_{R}$ and $A_{M}$ can be determined integrating over the
distributions once the lower and upper limits are specified. We
split the mass and radius intervals in a discrete number of bins,
and indicate with $n$ and $m$ the number of the mass and radius
bins, respectively. The normalized number
$N_{i}\left(M_{MC}\right)/N$ of clouds into the $i$ mass bin of $n$
bins and the number of clouds $N_{i}\left(R_{MC}\right)/N$ into the
$i$ radius bin of $m$ bins can be easily obtained. They are
independent of the normalization constant $A_{R}$ and $A_{M}$. The
sum over all the bins is of course normalized to unity. The fraction
of molecular clouds with mass belonging to the $i$-th bin of $n$
bins and radius belonging to the $j$-th bin of $m$ bins will be

\begin{equation}\label{fractionmr}
f\left(M_{i}, R_{j} \right)=\frac{N_{i}\left( M_{MC} \right)}{N}
\times \frac{N_{j}\left( R_{MC} \right)}{N}
\end{equation}

\noindent According to eqn. (\ref{tau_final}),  for $\eta=1000$ the
function $\Phi\left(\eta\right)\thickapprox 1$ and the central
density $\rho_{0,g}$ is nearly equal to the mean density because
dust is almost uniformly distributed, i.e.
$\rho_{0,g}\thickapprox\overline{\rho}_{g}$. Recalling that
$\overline{\rho}_{g}=3M_{MC}/\left(4 \pi R_{MC}^{3}\right)$, we
obtain $\tau_{\lambda} \varpropto M_{MC}/ R_{MC}^{2}$. This means
that for each ratio $M_{MC}/R_{MC}^{2}$ a different value of the
optical depth is expected.  Furthermore, splitting the range of
optical depths in $k$ bins, it may happen that different combinations
of $M_{MC}$ and $R_{MC}$ yield  a similar value of $\tau_{\lambda}$.
As a consequence of this, different pairs of bins $\left(i,j
\right)$ may give their fractional contribution to the same bin
$\tau\left(i,j\right)$ of the optical depth.

\begin{figure}
\psfig{file=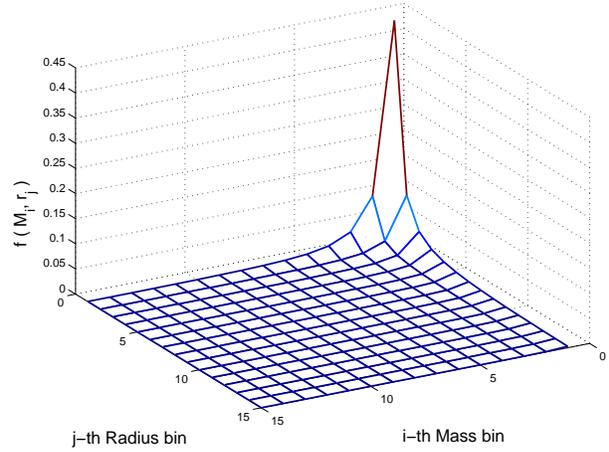,width=8truecm} \caption{Three
dimensional plot of the function $f\left(M_{i}, R_{j}\right)$ for
$15$ mass bins and $15$ radius bins. The first bin corresponds to
the lowest mass and radius of the distributions.}
\label{MassRadiusMC}
\end{figure}

The limits of the mass and radius distributions can be inferred from
the observational surveys of \citet{Solomon87} of galactic MCs and
the analysis made by \citet{Elmegreen96}. Typical values are:
$R_{up}\simeq 70$ pc, $R_{low}\simeq 5$pc, $M_{up}\simeq 3\times
10^{6} M_{\odot}$ and finally $M_{low}\simeq 4\times 10^{4}
M_{\odot}$. The exponents of the mass and  radius distribution,
$\alpha$ and $\beta$ respectively, are taken from
\citet{Elmegreen96}. We have $\beta\simeq 3.3\pm0.3$ for the
radii distribution and
$\alpha\simeq -1.8$ for the mass distribution. These power-laws
 imply that not all the optical depths
$\tau\left(i,j\right)$ will have the same weight, but those falling
in suitable $k$-bins will weigh more so that the whole range can be
reduced to these values. This is clearly shown in Fig.
\ref{MassRadiusMC} where the discrete function $f\left(M_{i},
R_{j}\right)$ is plotted over $15$ radius bins and $15$ mass bins
and the $f(1,1)$ bin weights much more than the other ones. However,
there is a strong dependence on the lower limits of the distribution
laws we have adopted, i.e. the one holding for the MW, and it is
likely that in different environments, such as strong star forming
galaxies, the distribution laws may be different. Therefore, a
realistic model of galaxies should include MCs with an ample range
of optical depths and also different laws for the mass and size of
the MCs. For the purposes of this study, we limit ourselves to the
distribution laws holding for the MW. With these distribution laws
and using 15 bins, the $f(1,1)$ bin weighs about $50 \%$ of the
total, as shown in Fig. \ref{MassRadiusMC}. Thanks to this, for the
time being we can simplify the problem and take only one optical
depth as a free parameter representative of the whole range, both in
star forming regions (Sect. \ref{compare_observations}) and galaxies
\citep{Piovan05}.

\section{SEDs of young  SSPs surrounded by MCs}\label{ised_dusty}

In this section, we present the SEDs of young SSPs in which the
effects of dust around are included according to the prescriptions
of Sect. \ref{rad_transport} and compare them with the standard case
in which dust is neglected. For each metallicity the analysis is
limited to the most probable bin $f(1,1)$ of the mass and radius
distribution of MCs. The optical depth of this bin is about
$\tau_{V}=35$ for the solar metallicity, $\tau_{V}=10$ and
$\tau_{V}=5$ for the LMC and SMC metallicities (extinction curves).
The scaling parameter for the MC dimensions is $R = 1$, while the
total carbon abundance per H nucleus in the log-normal populations
of very small carbonaceous grains, $b_{c}$, has been fixed to a
value slightly above the mean value of the allowed range. Finally,
the treatment of ionization is as in \citet{Weingartner01b}. More
details will be given in Sect. \ref{dusty_library} describing the
library of young SSPs with dust we have calculated.

\begin{figure}
\psfig{file=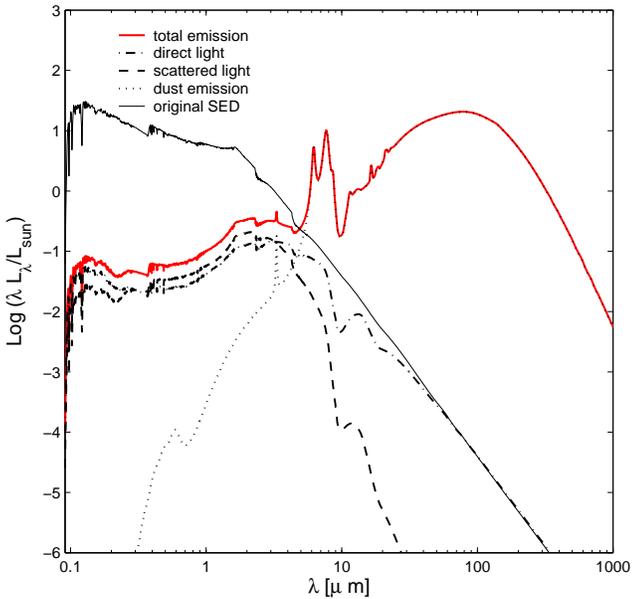,width=8.5truecm} \caption{SED of a SSP
of $30$ Myr and solar metallicity with and without dust. In the plot
we show the original spectrum of the SSP (thin solid line), the
total dusty spectrum (thick solid line), the contribution of direct
light (dot-dashed line), the contribution of scattered light (dashed
line) and, finally, the emission of dust (dotted line).}
\label{SSP30Myr}
\end{figure}

\begin{figure}
\psfig{file=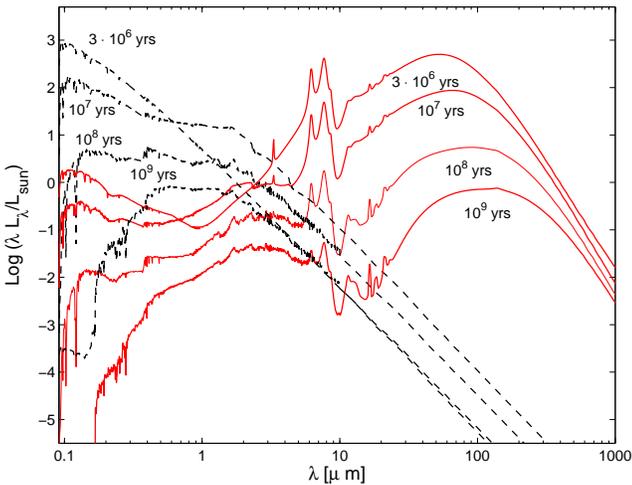,width=8.5truecm} \caption{SEDs of SSPs
of metallicity $Z=0.02$ without dust (dotted lines) and with dust
(solid lines). The spectra are obtained solving the radiative
transfer problem by means of the ray tracing method. The SSPs are
shown at four different ages, from top to bottom: $3$ Myr, $10$ Myr,
$100$ Myr and, finally, $1$ Gyr.} \label{tabul0.02}
\end{figure}

\begin{figure}
\psfig{file=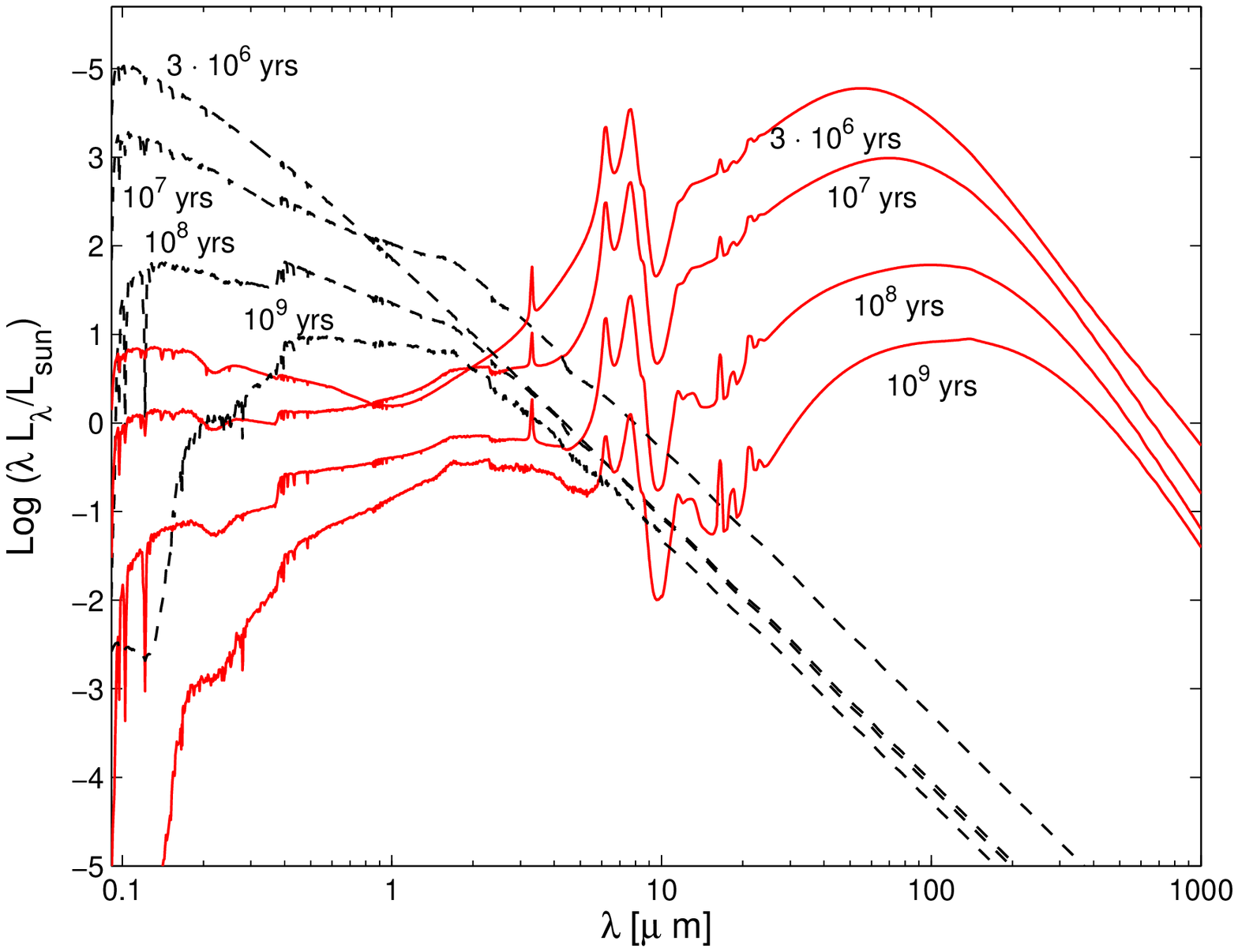,width=8.5truecm} \caption{The same as
Fig. \ref{tabul0.02} but for the metallicity $Z=0.008$.}
\label{tabul0.008}
\end{figure}

\begin{figure}
\psfig{file=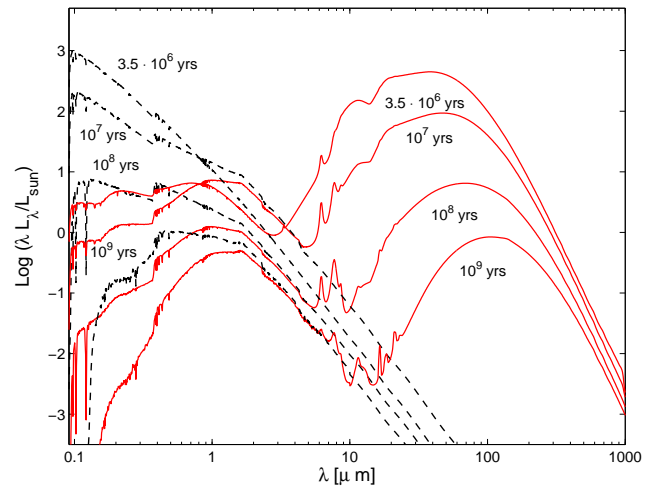,width=8.5truecm} \caption{The same as
Fig. \ref{tabul0.008} but for the metallicity $Z=0.004$ and the
youngest age that is $3.5$ Myr.} \label{tabul0.004}
\end{figure}

We begin showing in Fig. \ref{SSP30Myr} a SSP with $Z=0.02$ (solar
metallicity) and age of $30$ Myr, both in presence and absence of
dust. The main effects of dust are soon clear: the cloud in which
the young population is embedded absorbs the stellar radiation and
returns it in the IR. We may also note here the different
contributions by the various components of the specific intensity:
direct light, scattered light and dust emission. Passing to discuss
the effect of the metallicity, for the sake of brevity we will show
results only for $Z=0.02$ (Fig. \ref{tabul0.02}), $Z=0.008$ (Fig.
\ref{tabul0.008}), and $Z=0.004$ (Fig. \ref{tabul0.004}).

As expected, the SED shift toward the IR is stronger for solar or
LMC metallicities, because of the higher optical depth of the MCs
(all the other geometrical parameters of the cloud being kept
constant). For a given metallicity the peak in IR emission of cold
dust shifts toward shorter wavelengths at decreasing age of the SSP,
because the intensity of the radiation field gets higher. For each
shell of the radial grid in which a MC has been split, we calculate
the ionization state of PAHs. The \citet{Weingartner01b} model
yields SEDs of very young SSPs in which the PAHs are almost fully
ionized and the profile of aromatic infrared bands of PAHs (AIBs) is
very close to that for the ionized cross sections, as shown by
Figs. \ref{tabul0.02}, \ref{tabul0.008} and \ref{tabul0.004}. At
increasing age and hence decreasing radiation field, the profile of
AIBs shifts to that of neutral cross section because more and more
PAHs become neutral. For the lowest metallicities the contribution
of PAHs becomes small because of the small percentage of
carbonaceous grains in the adopted extinction curve (the one for
SMC). Furthermore, at decreasing age, AIBs become weaker due to the
combined effects of the lower optical depth and higher radiation
field which cause emission by hot dust in the same wavelength
interval of AIBs.

\subsection{A library of young SSPs surrounded by MCs }\label{dusty_library}

Dealing with theoretical spectra of young SSPs embedded into dusty
MCs is a cumbersome affair as a large number of parameters is
involved. First, we have those related to dust and its composition
(still uncertain), together with those of the optical properties and
distribution laws of the grains. Second, we have the geometrical
parameters introduced to solve the radiative transfer in a medium
with sources distributed in it. Although one could simply take a
sort of prototype MC to be inserted in population synthesis studies,
it is more physically grounded and close to reality to create and
explore in some detail the space of parameters.

First of all, let us discuss the optical depth $\tau_{\lambda}$ and
the scaling factor $R$ of MCs (the latter fixing their size). For
more details about $R$ and $\tau_{\lambda}$ see also
\citet{Takagi03}.

\textit{$\tau_{\lambda}$}: the optical depth, see eqn.
(\ref{tau_final}), is the main parameter of the radiative transfer
problem. Its effect on the spectrum emitted by the MC is simple to
describe: if the optical depth is high, more energy is shifted
toward longer wavelengths. This amount of energy first quickly
increases at increasing optical depth, then becomes less sensitive
to $\tau_{\lambda}$ and even tends to flatten out for high
$\tau_{\lambda}$. This can be seen in Figs. \ref{tabul0.02} and
\ref{tabul0.008}, where even if the optical depth is different, yet
it is high enough to show the above effect. The optical depth
depends on the mass and radius of the MC (the geometrical
parameters) and different distribution laws for masses and radii
lead to different optical depths. The ideal case would be to set up
a library of SSPs covering an ample range of optical depths. Because
of the high computational time required, the present release of the
library is limited to two values of the optical depth for each
metallicity, namely  $\tau_{V} = 35$ and $\tau_{V} = 5$.

\textit{$R$}: the scaling parameter $R$ has been introduced because
the mass $M_{\star}$ contained inside a MC has been normalized to
$1\, M_{\odot}$, in view of the use of these SSPs in evolutionary
population synthesis models for galaxies. $R$ links the mass of the
sources of radiation to the dimension of the cloud, as described in
Sect. \ref{opt_depth}. There is no effect of varying $R$ over the
part of the SED given from the absorbed stellar emission. The only
effect of $R$ is to change the position of the FIR peak due to dust
emission. A ideal MC scaled with a larger value of $R$ will have a
lower temperature profile of the grains because of the bigger
dimensions. Accordingly, the FIR peak of dust emission will simply
shift to longer wavelengths. This effect is entangled with the age
effect, because for older populations the radiation field will be
weaker and the FIR peak will undergo a similar shift to longer
wavelengths.

Let us now examine the parameters controlling the MIR region of the
spectrum, i.e.  the intensity of PAH emission and the AIB
profiles.

\begin{figure}
\psfig{file=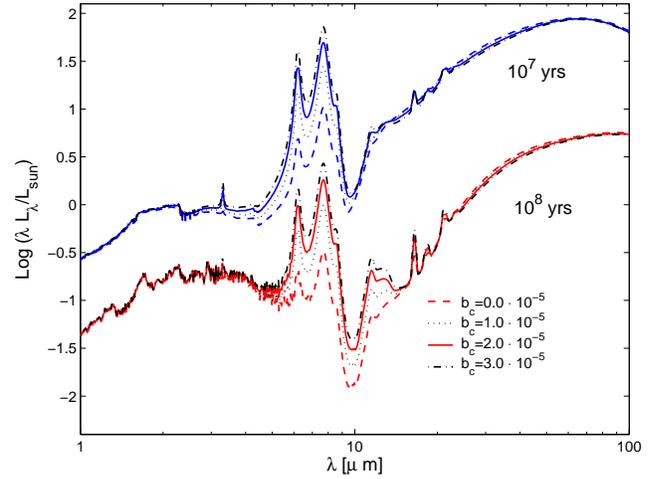,width=8.5truecm} \caption{PAH emission
as a function of the parameter $b_{C}$. The MIR-FIR region is shown
for two young dusty SSPs of $10^{7}$ and $10^{8}$ Myr and
metallicity $Z=0.02$. Four values of $b_{C}$ are taken into account,
going from $0 \cdot 10^{-5}$ until $3 \cdot 10^{-5}$, the upper
limit for the $R_{V}=5.5$ extinction curve \citep{Weingartner01a}.
The other parameters are kept fixed, $R = 1$, $\tau = 35$ and the
ionization model of \citet{Weingartner01b}.} \label{BcZ02Rv5e5}
\end{figure}

\begin{figure}
\psfig{file=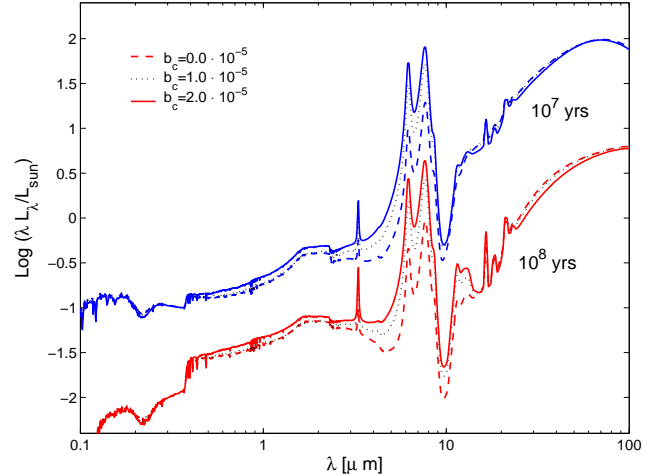,width=8.5truecm} \caption{PAH emission
in function of the parameter $b_{C}$. The MIR-FIR region is shown
for two young dusty SSPs of $10^{7}$ and $10^{8}$ Myr and
metallicity $Z=0.008$. Three values of $b_{C}$ are taken into
account, going from $0 \cdot 10^{-5}$ until $2 \cdot 10^{-5}$, the
upper limit for the average LMC extinction curve, as proposed in
\citet{Weingartner01a}. The other parameters are kept fixed, $R =
1$, $\tau = 35$ and ionization model of \citet{Weingartner01b}.}
 \label{BcZ008LMC}
\end{figure}

\textit{$b_{C}$}: this parameter is the total C abundance (per H
nucleus) in the two log-normal distribution laws of very small
carbonaceous grains. Even if other distribution laws for the grain
dimensions can be used, our preference goes to the
\citet{Weingartner01a} law because it has been proved to provide
excellent fits to the extinction curves and it contains the
contribution of very small grains to carbon abundance as a free
parameter. Only an upper limit to the parameter $b_{C}$ can be
derived and all the values lower than or about equal to this upper
limit are possible (see Sect. \ref{dust_model}). However the exact
value of $b_C$ bears very much the PAH emission in the MIR. At
increasing $b_{C}$, the AIBs are more pronounced with higher flux
levels, while the flux in the UV-optical region and the shape of the
absorbed stellar emission remain unchanged as the extinction curve
is exactly the same for different values of $b_{C}$. At decreasing
$b_{C}$, the emission in the FIR slightly increases (the total
energy budget has clearly to be conserved) because the global
abundance of $C$ is fixed, and the distribution of the grains has to
compensate for the small number of VSGs with respect to the higher
number of BGs. In Figs. \ref{BcZ02Rv5e5} and \ref{BcZ008LMC} we show
the effect of varying $b_{C}$ on the emission of two young dusty
SSPs of $10^{7}$ and $10^{8}$ Myr both for the metallicities $Z =
0.02$ and $Z = 0.008$. At increasing $b_{C}$, the AIB emission
increases, whereas the FIR emission goes in the opposite way. In our
library, four values of $b_{C}$ for solar or super-solar
metallicities are considered, three values are taken into account
for the LMC metallicity, and only one value is explored for the SMC
and lower metallicities \citep[see also][]{Weingartner01a}.

\textit{PAH ionization}: together with $b_{C}$, there is another
parameter affecting the emission of PAHs to be taken into account,
in order to match also the complicate profiles of the AIBs, i.e. the
ionization state of the PAHs. The profile of the bands depends
indeed on the PAH ionization, because the optical properties are
different for ionized and neutral PAHs as been discussed in Sect.
\ref{casting}.
The ionization is calculated for each spherical shell of a MC. It
strongly depends on the age of the SSP: passing from profiles of
ionized PAHs for the youngest populations (because of the strong
radiation field) to those of neutral PAHs at old ages (because  of
the weaker radiation field).

\begin{figure}
\psfig{file=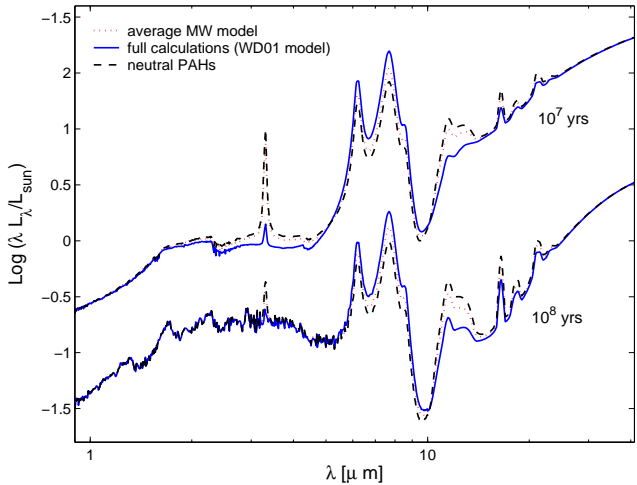,width=8.5truecm} \caption{PAH emission
as a function of the ionization model. The MIR-FIR region is shown
for two young dusty SSPs of $10^{7}$ and $10^{8}$ Myr and
metallicity $Z = 0.02$. Three models of ionization are taken into
account as indicated.} \label{IonizzazioneZ02}
\end{figure}

In order to fully explore the space of parameters, three ionization
models are considered. The first  one  is by \citet{Weingartner01b}.
The second model strictly follows the same ionization profile
calculated in \citet{Li01} for the diffuse ISM of the MW. The third
model simply takes into account  only the optical properties
of neutral PAHs. In this way we can study a sequence of ionization
states going from strongly ionized to almost neutral. For the second
and the third model there is no change of the profile of the AIBs at
varying the age of the underlying SSP as instead it happens in the
first model. In Fig. \ref{IonizzazioneZ02} we compare the results
for the three ionization models showing the emission of two young
dusty SSPs with different age, i.e. $10^{7}$ and $10^{8}$ Myr, and
the same metallicity $Z = 0.02$ (all other parameters are kept
fixed). The PAH ionization has no effect on the UV-optical and FIR
emission because the distribution of the grains is the same and
their optical properties in these wavelength intervals do not depend
on the ionization state. In the MIR we can notice that, as expected,
the models with more ionization have more enhanced AIBs at $6,2$ and
$7.7\,\mu m$ and $8.6\mu m$, and weaker AIBs at $3.3\,\mu m$ and
$11.5 \,\mu m$. Clearly, the neutral model, has the strongest AIB at
the $3.3 \,\mu m$.

\subsection{Evaporation of the MCs} \label{evaporation_MCs}

\begin{figure}
\psfig{file=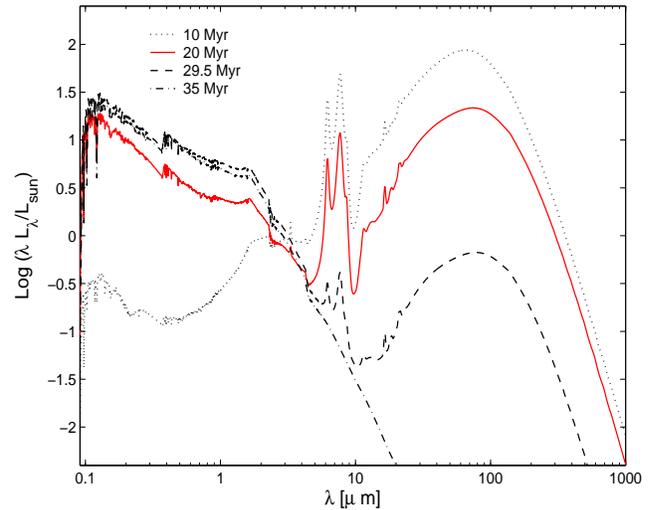,width=8.5truecm,height=6.8truecm}
\caption{SEDs of a SSP of metallicity $Z=0.02$ from the age of $10$
Myr (dashed line) to the age of $35$ Myr (dot-dashed line). Two
intermediate ages are shown of $20$ Myr (continuous line) and $29.5$
Myr (dashed line), respectively. The parameter $t_{0}$ is set equal
to 15 Myr.} \label{newevol0.02}
\end{figure}

\begin{figure}
\psfig{file=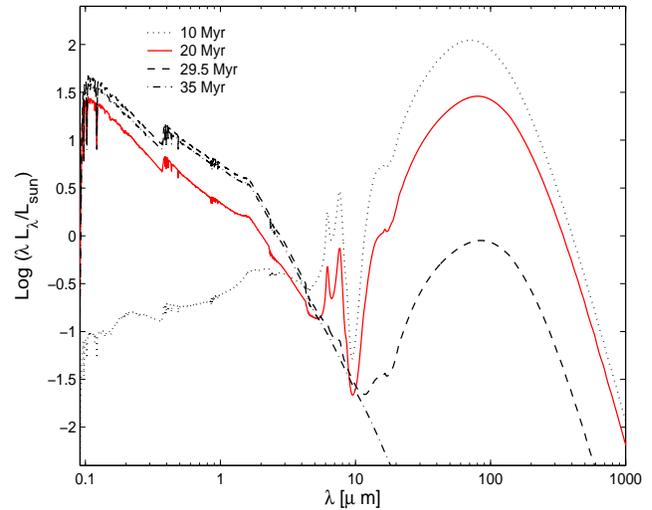,width=8.5truecm,height=6.8truecm}
\caption{The same as Fig. \ref{newevol0.02} but for the metallicity
$Z=0.004$.} \label{newevol0.004}
\end{figure}

Before using the SED of young dusty SSPs to derive the SED of
galaxies, an important aspect has to be clarified. How long does it
take to evaporate the MC surrounding a new generation of stars? The
subject must be clarified in advance, because what we have been
doing so far is only to take the energy emitted by a SSP and inject
it into a MC to be re-processed by dust: we have not specified the
evolution of the MCs, which are most likely to dissolve by the
energy input of the stars underneath. Following \citet{Silva98}, the
process can be approximated by gradually lowering as function of
time the amount of SSP flux that is reprocessed by dust and
increasing the amount of energy that freely escapes from the MC.
This is, of course, a rough approximation of the real situation in
which the energy input from supernovae and massive stars gradually
sweep away the parental gas and dust. A more realistic approach
should take into account the energy input coming from supernovae
explosions and massive stars and compare it to the binding energy of
the gas.

Let us denote with  $f_{DSSP}$ the fraction of the SSP luminosity
that is reprocessed by dust and with $t_{0}$ the time scale for
a MC to evaporate

\begin{equation}
f_{DSSP}=\left\{
\begin{array}{llll}
1          & & &\quad t\leq t_{0} \\
2-t/t_{0}  & & &\quad t_{0}<t\leq 2t_{0} \\
0          & & &\quad t\geq t_{0}
\end{array}
\right.
\end{equation}

\noindent Accordingly, the fraction of SSP luminosity that escapes
without interacting with dust is $f_{SSP}=1-f_{DSSP}$.

\noindent The parameter $t_{0}$ will likely depend on the properties
of the ISM and type of galaxy in turn. Plausibly, $t_0$ will be of
the order of the lifetime of massive stars, say from 100 to 10
$M_\odot$, that first  explode as supernovae and whose ages range
from $3$ to $100$ Myr. However, the real timescale is likely to be
much close to the shortest value, i.e. the lifetime of the most
massive supernovae in the SSP, for a low density environment like
spiral galaxies with a low star formation rate, while it will be
much close to the longest value for a high density environment like
the star forming central regions of starburst galaxies. For the sake
of illustration, in Figs. \ref{newevol0.02} and \ref{newevol0.004}
we show the gradual transition of SSPs with different metallicity,
$Z = 0.02$ and $Z = 0.04$, respectively, from fully embedded in
their original MCs to free of dust. In all the cases, the optical
depth is  $\tau_{V} = 35$ and $R = 1$, $b_{c}$ is fixed at the high
value holding for MW and at the only available value for the SMC
and, finally, the \citet{Weingartner01b} ionization model is fully
used. The SEDs are displayed at four different ages, $10$, $20$,
$29.5$, and $35$ Myr. The parameter $t_{0}$ is fixed to $t_{0} = 15$
Myr.

At the age of 10 Myr, the SSPs are fully enshrouded by dust and no
evolution of the fluxes is still simulated (dotted lines in Figs.
\ref{newevol0.02}, and \ref{newevol0.004}). At the age of 35 Myr,
the SSPs are no longer surrounded by MCs, their SED is the one of a
bare classical SSP without dust contribution of extinction and
emission (dot-dashed lines in Figs. \ref{newevol0.02}, and
\ref{newevol0.004}). For the other two values of the age the SED is
intermediate to the previous ones. The trend is as follows: at the
beginning the stars are heavily masked but when the age of the SSP
enter the range between $t_{0}\leq t \leq 2t_{0}$ part of the energy
is transferred to the optical range and part remains in the IR. As
the stars become older, the quantity of energy emitted in the IR
diminishes and eventually disappears when $t\geq 2t_{0}$.

\section{Comparison with observational data}
\label{compare_observations}

To test our model for dusty SSP we compare the theoretical SEDs with
the observational data for the central star forming regions of some
local starburst galaxies. A simple  picture aimed at explaining most
of the observational constraints for starbursters has been proposed
by \citet{Calzetti01}. The newly born stars, likely still embedded
in their parental MCs, are concentrated in the centre of the galaxy,
surrounded by long-lived less massive stars wrapped by a diffuse
dusty medium. Clearly, the real distribution of stars and dust will
be much more complicated, with clumps and asymmetries. If the burst
of star formation is strong enough, the flux coming from  the
central region of the galaxy (small-aperture data) is dominated by
the obscured light coming from the new generations of stars. In this
way, if we limit ourselves to study small central regions of
starbursters, to a first approximation we can try to reproduce the
observed SED with an obscured SSP of suitable mass. It is clear,
however, that if the burst lasts longer  than the life of the most
massive stars \citep{Calzetti01}, the stellar populations in the
central region under examination are no longer coeval. Rather they
are a mix of spatially concentrated very young stars superimposed to
older and more dispersed foreground stars. The longer or the weaker
the burst of star formation, the higher is the influence of the
older generations of stars on the resulting SED and the weaker the
physical significance of the SSP approximation. Nevertheless,
despite of these remarks, the comparison of our SSP SEDs with the
SED from the central dusty star forming regions of local
starbursters is the best tool to our disposal to validate the model.

\subsection{Central region of Arp 220 } \label{Arp_220}

Arp $220$ is a very bright object in the Local Universe, classified
by \citet{Arp66} as a peculiar galaxy. The interest in this galaxy
became very strong when the IRAS satellite revealed that it is a
powerful source of FIR emission. Today, we know that Arp $220$ is
the nearest and most thoroughly studied example of UltraLuminous
InfraRed Galaxies (ULIRG), objects with an infrared luminosity
$L_{IR} \geq 10^{12} L_{\odot}$ \citep{Sanders88}. The redshift of
the galaxy, taken from the NED database is $z = 0.01813$, which  for
the Hubble constant $H_{0} = 72$ km/s/Mpc yields  a distance of
about $76$ Mpc. This value is fully consistent with the distances
proposed in \citet{Soifer87} and in \citet{Spoon04}.
\citet{Wynn-Williams93} showed that almost all the MIR flux of
Arp$220$ comes from a small central region of about $5^{"}$
aperture.  This concentration of the MIR emission has been confirmed
by \citet{Soifer99} comparing the fluxes measured varying the beam
size at a fixed MIR wavelength. We take the $5^{"}$ central region
of the galaxy, which corresponds to a diameter of about $2$ kpc
\citep{Takagi03}, to test our SSPs.

\begin{figure*}
\centerline{ \psfig{file=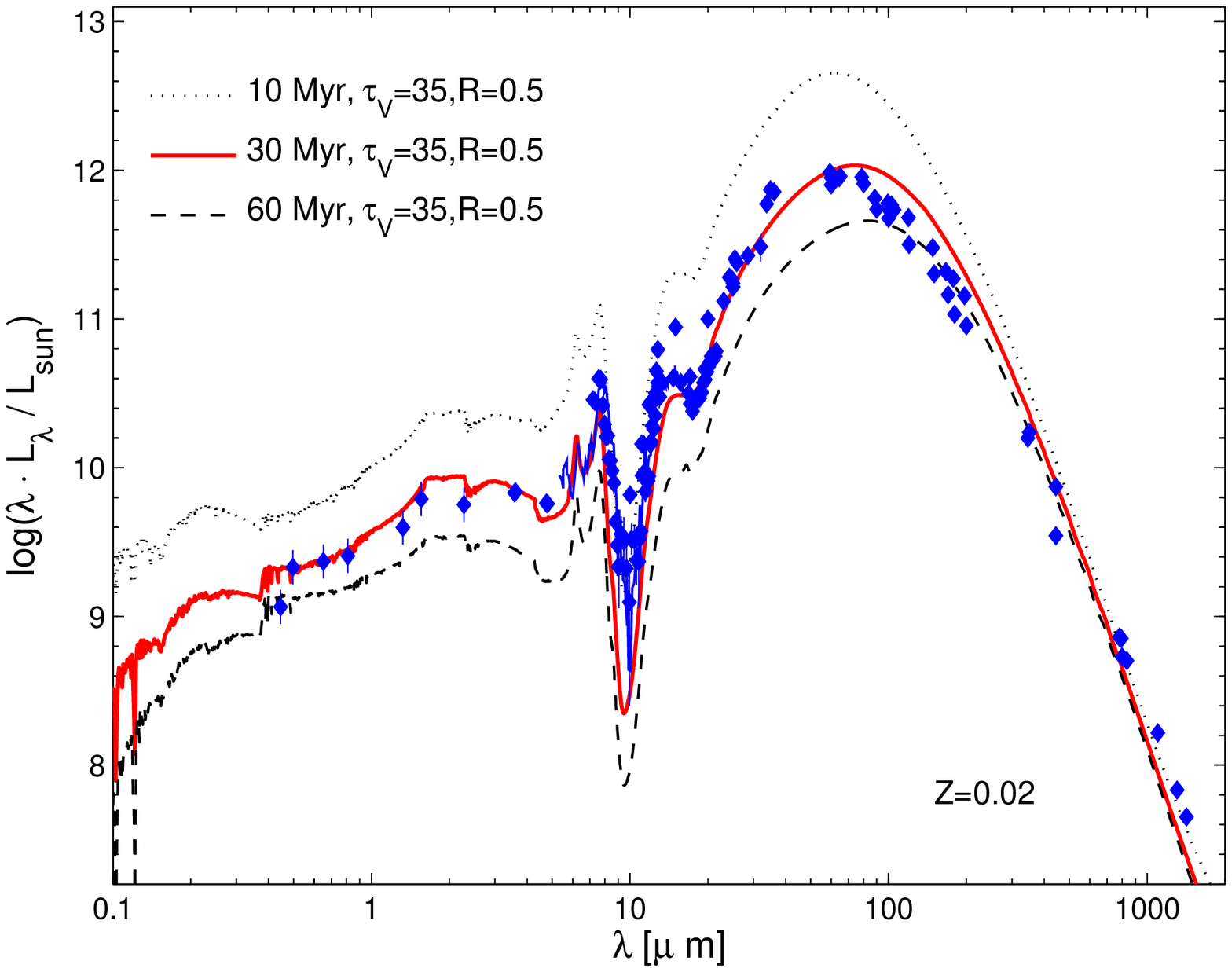,width=8.2truecm}
\psfig{file=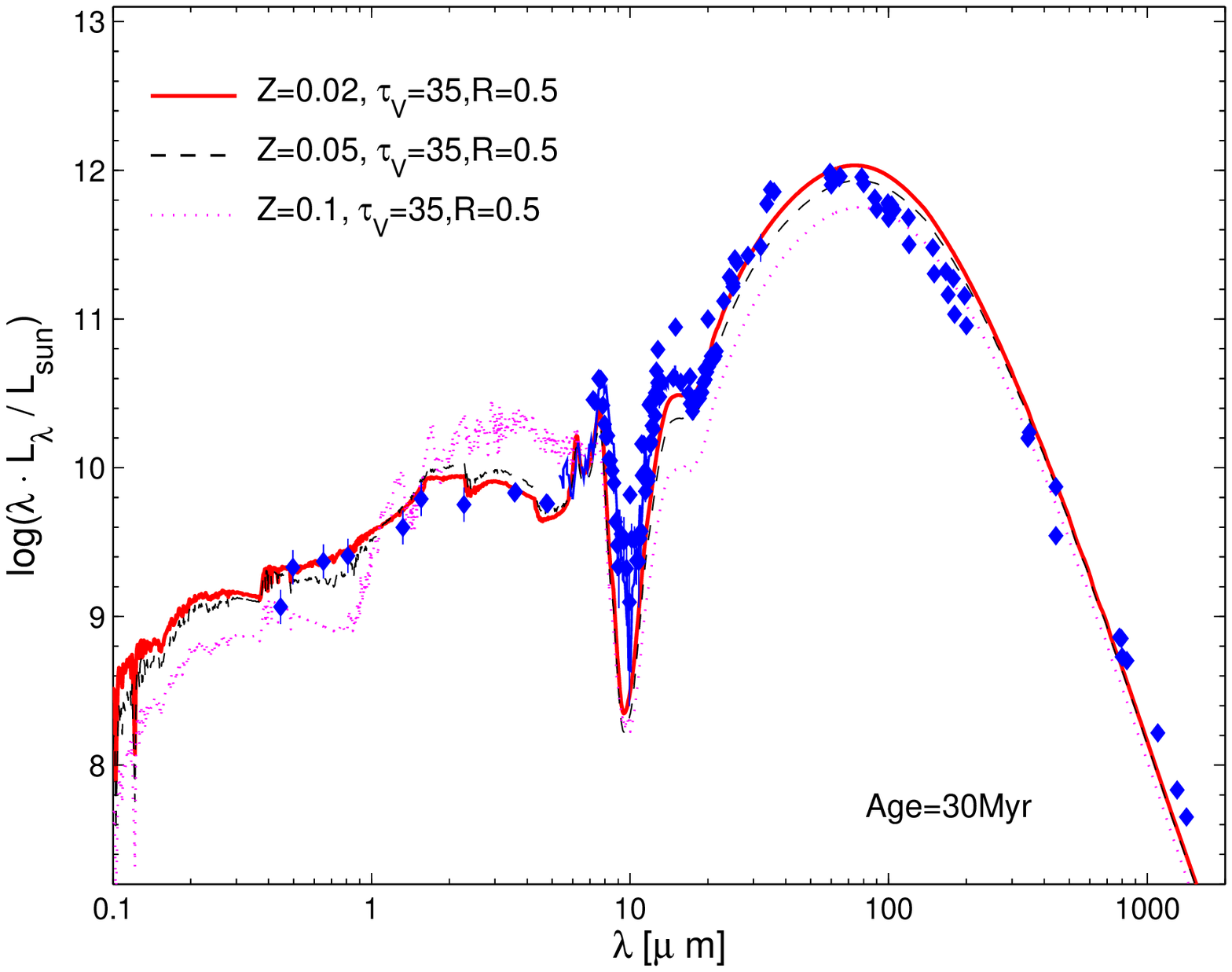,width=8.2truecm}} \caption{{\bf Left
Panel}: Best fit of the emission  from the central region of
Arp$220$. We show three SSPs with the ages of $10$, $30$ and $60$
Myr (dotted, continuous and dashed lines respectively). The   best
fit is for $30$ Myr. Data are taken from \citet[][B, g, r,
i]{Sanders88}, \citet[][J, H, K, L]{Carico90}. The MIR fluxes are
from \citet[MIR]{Spinoglio95}, \citet[MIR]{Smith89}, \citet[from MIR
to FIR]{Klaas97}, \citet[MIR]{Wynn-Williams93}, \citet[5 - 16 $\mu$m
ISOCAM-CVF spectrum]{Tran01}. FIR and radio data are from
\citet[sub-mm]{Rigopoulou96}, \citet[UKIRT sub-mm]{Eales89},
\citet[SCUBA sub-mm]{Dunne00}, \citet[SCUBA sub-mm]{Dunne01}. Other
data are taken from \citet{Spoon04}. {\bf Right Panel}: The same as
in the left panel but for  fixed age and varying metallicity. We
show three metallicities: $Z=0.02$, $Z=0.05$ and $Z=0.1$
(continuous, dashed and dotted lines, respectively).}
\label{ARP220SSP}
\end{figure*}

In Fig. \ref{ARP220SSP} we show the results of our fit obtained
using dusty SSPs, with high optical depth $\tau_{V}=35$, compact
structure with $R=0.5$ and the SMC extinction curve. The low
metallicity extinction curve we have selected agrees with the
results by \citet{Takagi03}. Only with a mixture of grains, like the
one of the SMC extinction curve, we were able to reproduce the deep
absorption features of silicates, the low emission of PAHs and the
optical/NIR continuum. So, the fit seems to require relative
proportions of grains favouring the grains of silicates with respect
to the graphitic ones, like in a flat SMC extinction curve, even if
the average metallicity of the environment could be easily
supersolar. To clarify the issue we should apply the EPS to chemical
models in which the dust abundances \citep{Dwek98,Dwek05} are
followed in detail. As this step is not yet accomplished, work is in
progress. The best fit is found for the age of $30$ Myr, a mass of
$0.4\cdot 10^{11}M_{\odot}$ and both solar and supersolar
metallicity, thus suggesting the present of a high metallicity
environment as shown in the right panel of Fig. \ref{ARP220SSP}.

\subsection{Central region of NGC 253} \label{NGC_253}

NGC 253 is a bright starburst galaxy belonging to the Sculptor
group, in the direction of the South Galactic Pole. It is perhaps
one of the nearest Galaxies besides those of the Local Group. The
redshift is $z=0.000804$ (NED), corresponding  to the distance
$d=3.35$Mpc for $H_{0} = 72$ km/s/Mpc.

The central region we have selected is within $15"$ aperture and its
SED is centered around the peak of MIR emission. For this region
there are available data on the PAH emission from
\citet[][ISO-SWS]{Sturm00} and \citet[][ISOCAM]{ForsterSchreiber03}
that agree each other, thus confirming the accuracy of the
calibrations and setting a strong constraint on the global fit.
There are some problems with OAO data \citep{Code82} covering the UV
range, because of the large aperture $10^{'}$. We scaled the OAO
data to the aperture size in usage here. It is clear that a precise
correction would require the knowledge of the true UV profile of the
galaxy, which is unknown. For this reason, even if the corrected UV
level of flux is reproduced, we will give less weight to the  UV
region looking for the global fit.

\begin{figure*}
\centerline{ \psfig{file=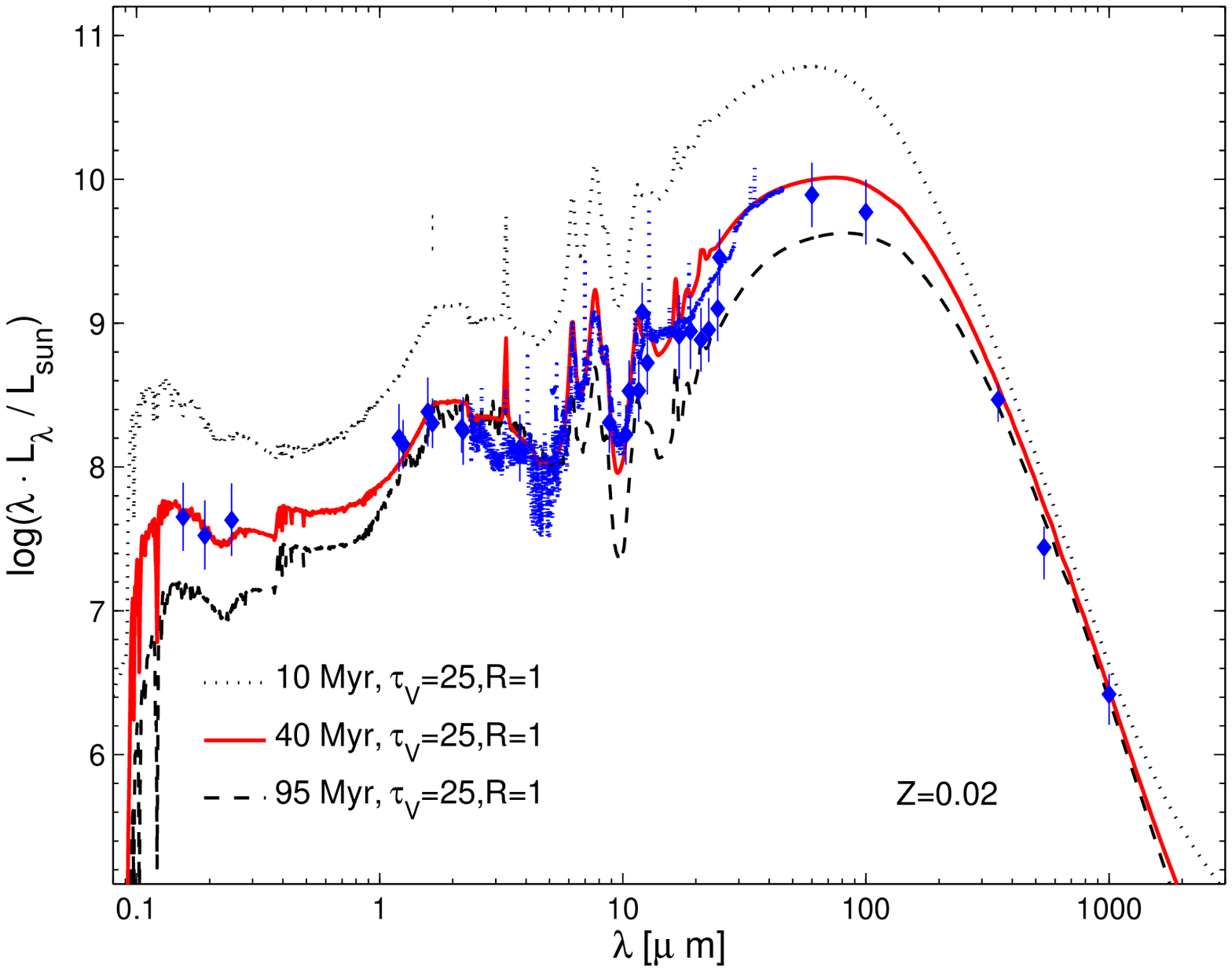,width=8.2truecm}
\psfig{file=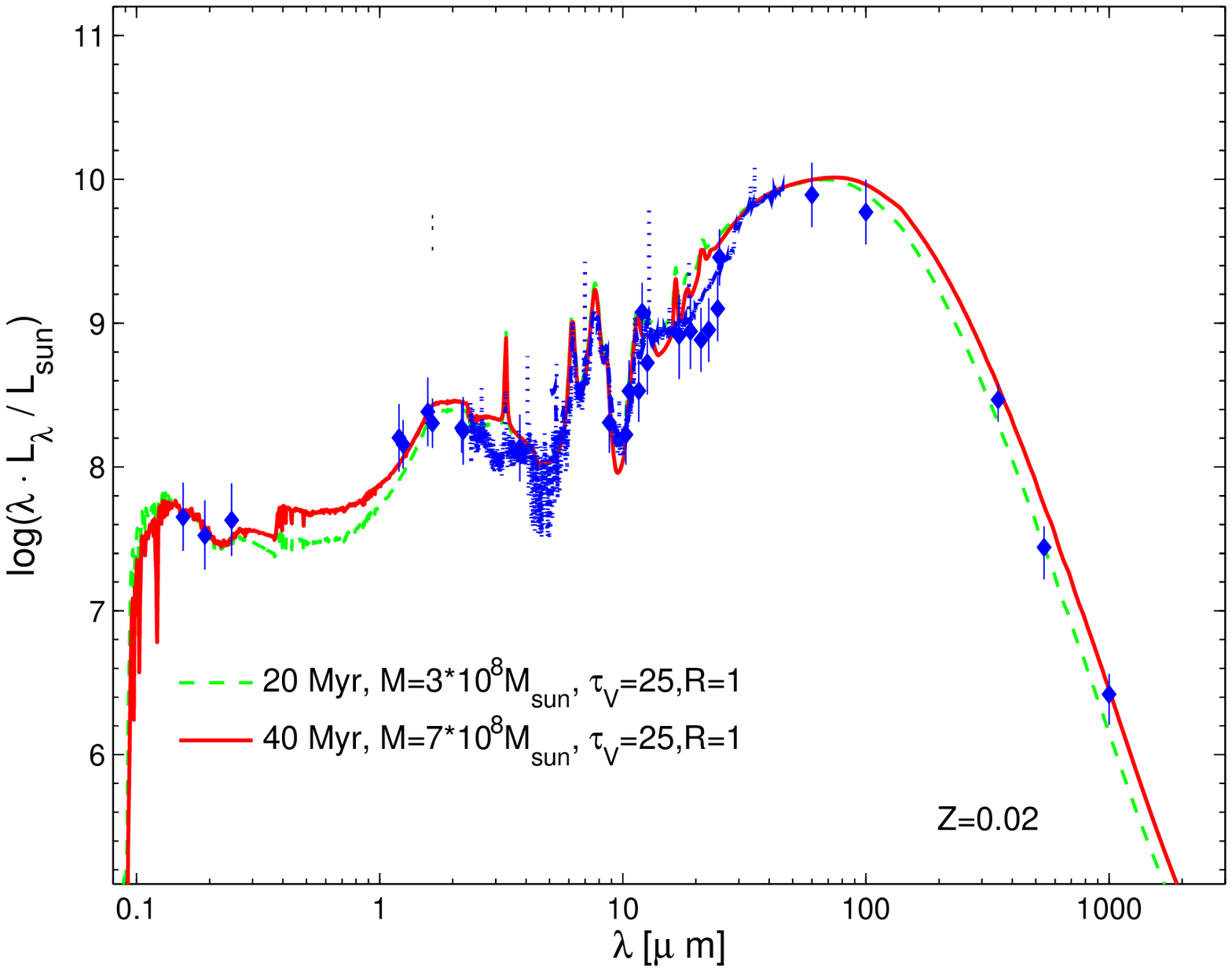,width=8.2truecm}} \caption{{\bf Left
Panel}: Best fit of the emission from the central region of
NGC$253$. We show three SSPs with ages of $10$, $40$ and $95$ Myr
(dotted, continuous and dashed lines, respectively). The best fit is
for $40$ Myr. Data for the central region are taken and adapted from
\citet[UV]{Code82}, \citet[J, H, K, L]{Spinoglio95}, \citet[J, H, K
- 2MASS]{Jarrett00}, \citet[IRAS]{Moshir90},
\citet[sub-mm]{Chini84}, \citet[][ private communication]{Sturm00},
\citet[][ private communication]{ForsterSchreiber03}. {\bf Right
Panel}: The same as in the left panel but at varying both the age
and the mass of the galaxy for the same metallicity.}
\label{NGC253SSP}
\end{figure*}

In the right panel of Fig. \ref{NGC253SSP}, we show our best fit
obtained with a SSP of solar metallicity, $\tau_{V}=25$, $R=1$, MW
extinction curve for dense regions $(R_{V}=5.5)$ and average-low
value of the abundance of carbon in the very small carbonaceous
grains $(b_{C}=1.0 \cdot 10^{-5})$. The ionization model which best
agrees with the observations is the one with a dominant neutral
component of PAHs. The mass of the population of stars is about $7
\cdot 10^{8} M_{\odot}$. In the left panel we show an effect of a
sort of age-mass degeneracy: if we can obtain a good fit for a given
age and mass, an equally good fit is possible lowering both the age
and the mass, because a younger population is more luminous. In
spite of this uncertainty, there should be a young population of
stars with age of about $30$ Myr and total mass of about $5 \cdot
10^{8} M_{\odot}$.

\subsection{Central region of M82 } \label{M_82}

The starburst galaxy M$82$ with its companion M$81$ was discovered
in 1774 by Johann Bode and independently by M\'{e}chain in 1779. He
reported them to Messier, who added these two objects to his
catalog. The central region of the galaxy, which likely suffered a
strong interaction about $10^{8}$ years ago
\citep{ForsterSchreiber03} with its companion M$81$, shows a
remarkable burst of star formation activity. The classical distance
to M$82$ is $3.25 \pm 0.20$ Mpc \citep{Tammann68} that makes of
M$82$ the nearest and most studied starburst galaxy \citep[see][and
references therein]{Shopbell98}.

For M$82$ there are many available data for the PAH emission in the
MIR region, to which we pay particular attention. They are very
useful to constrain the fit to the observed SED of the central
region. ISO-SWS data by \citet[][private communication]{Sturm00}
cover a wide spectral region going from $2.4$ to $45 \mu m$.
Different SWS full grating scans that are observed with different
aperture sizes, going from $14^{"} \times 20^{"}$ to $20^{"}\times
33^{"}$ are patched together. Recently, \citet[][ private
communication]{ForsterSchreiber03} observed M$82$ in the MIR range
between $5.0$ and $16 \mu m$ with ISOCAM on board of ISO. The total
field of view was $96^{"}\times 96^{"}$. We have chosen to examine
the central M$82$ region of $30^{"}\times 30^{"}$. For this coverage
we can use the \citet{ForsterSchreiber03}  data (Core plus Disk
observations), while \citet{Sturm00} data with smaller apertures
have been properly corrected.

\begin{figure}
\psfig{file=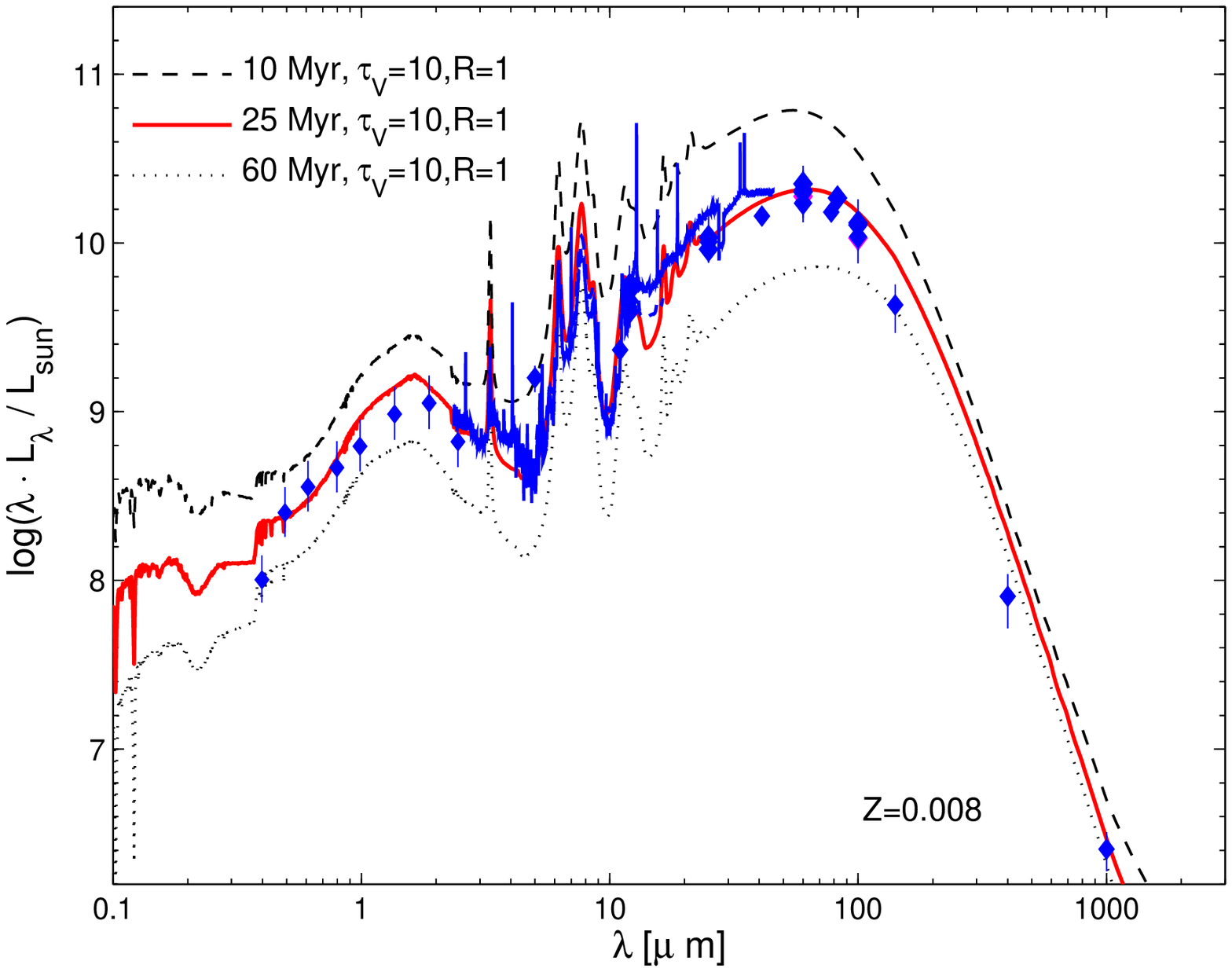,width=8.2truecm} \caption{Modelling the
SED of the $30^{"}\times 30^{"}$ central region of M$82$. We show
three SSPs with  ages of $10$, $25$ and $60$ Myr (dashed, continuous
and dotted lines, respectively). The best fit is for $25$ Myr. Data
for the central region of this galaxy are taken from: NED online
database, \citet[UBVRI]{Johnson66}, \citet[JHK]{Ichikawa95},
\citet[IRAS]{Soifer87}, \citet[IRAS]{Golombek88},
\citet[IRAS]{Rice88}, \citet[FIR and sub-mm]{Klein88},
\citet[][private communication]{Sturm00}, \citet[][private
communication]{ForsterSchreiber03}.} \label{M82SSP}
\end{figure}

In Fig. \ref{M82SSP} we show our best fit, obtained with
$\tau_{V}=10$, $R=1$, metallicity $Z=0.008$ and the LMC extinction
curve. The carbon abundance in small grains is equal to the average
$b_{C}=1.0 \cdot 10^{-5}$ of current estimates for LMC. The best
ionization state of PAHs is the average ionization model for the MW
diffuse ISM. All the spectral regions are well reproduced, the major
uncertainties being the $18 \mu m$ silicates absorption feature and
the far UV region where we have no data. The mass obtained from the
fit is $M=8 \cdot 10^{8}M_{\odot}$.

\subsection{Central region of NGC 1808 } \label{NGC_1808}

NGC 1808 is a starburst spiral galaxy in the southern hemisphere
sky. The redshift of the galaxy is $z = 0.003319$ (NED) which for
$H_{0} = 72$ km/s/Mpc corresponds to a distance of about $13.82$
Mpc.

\begin{figure}
\psfig{file=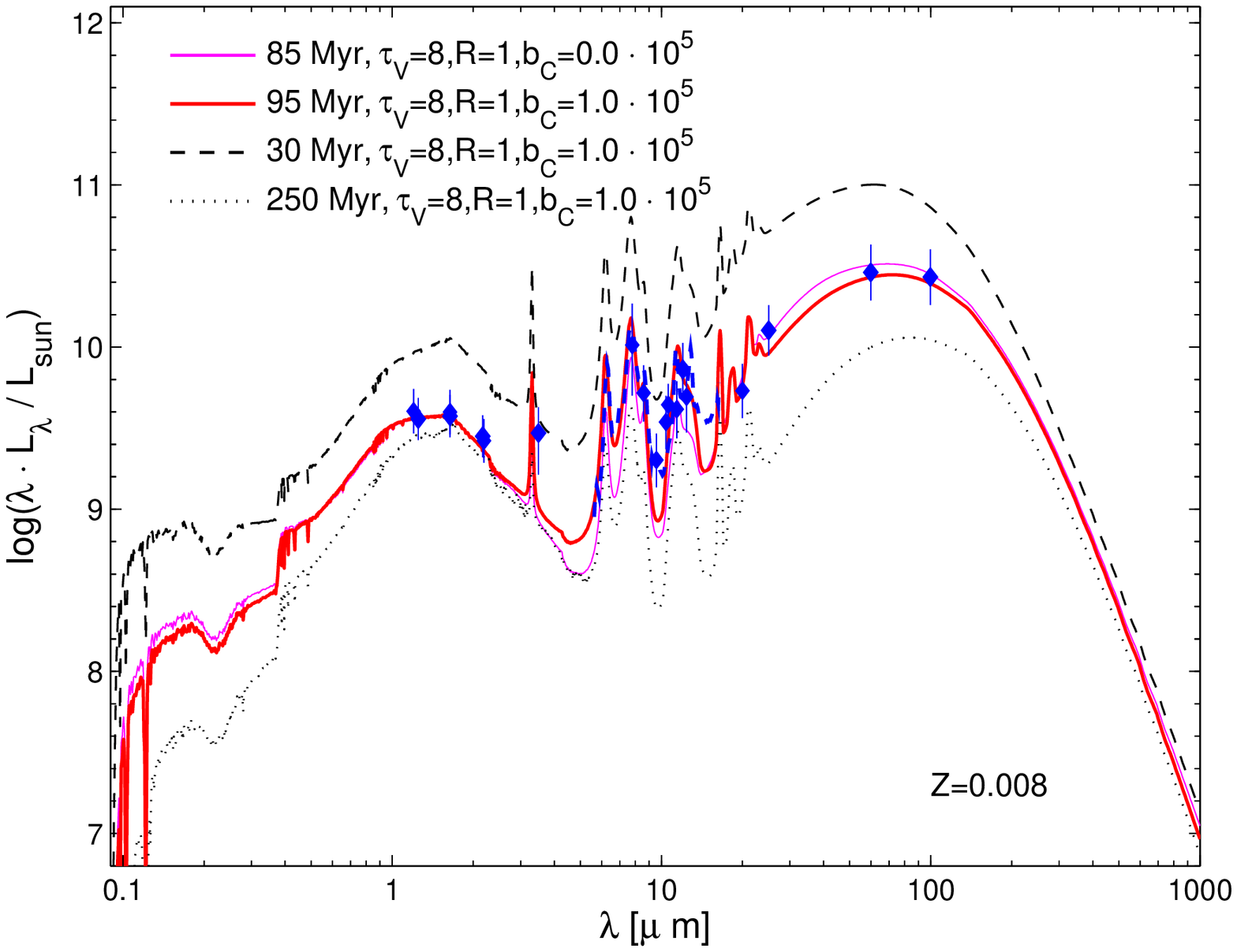,width=8.2truecm} \caption{Modelling the
SED of the $15^{"}\times 15^{"}$ central region of NGC$1808$. We
show three SSPs with  ages of $30$, $95$ and $250$ Myr (dashed,
thick continuous and dotted lines, respectively). The the best fit
is for $95$ Myr for $b_{C}=1.0 \cdot 10^{5}$. Another good fit (thin
continuous line) with different $b_{C}$ is also shown. Data for this
galaxy are taken in the NED online database from
\citet[JHK]{Griersmith82}, \citet[JHK]{Jarrett00},
\citet[HKL]{Glass73}, \citet[MIR, 8-13 $\mu$m]{Frogel82},
\citep[MIR][private communication]{ForsterSchreiber03} and
\citet[IRAS]{Moshir90}.} \label{NGC1808SSP}
\end{figure}

For NGC 1808 we have chosen the region $15^{"}\times 15^{"}$ in the
centre of the galaxy. In Fig. \ref{NGC1808SSP} we show four models,
all calculated for the metallicity $Z=0.008$, optical depth
$\tau_{V}=8$, scale radius $R=1$. The best fit is obtained for the
age of $95$ Myr and a carbon abundance in the log-normal populations
of VSGs of $b_{C}=1.0 \cdot 10^{-5}$. A good fit is also possible
for a lower abundance $b_{C}=0$ and slightly different age of $85$
Myr (see Fig. \ref{NGC1808SSP}). In all models, the total stellar
mass is $M=5 \cdot 10^{9}M_{\odot}$. The case of NGC 1808 is more
problematic than the other galaxies we have examined insofar. First,
there are no data  for the optical region with the desired aperture
and the aperture-correction applied to other optical data published
in literature is too problematic. Second it is difficult to
simultaneously reproduce both the flux level of the PAH emission and
the absorption features of silicates at $9.7 \mu m$ and $18 \mu m$.
In our fits the absorption features are too deep. The ideal solution
would require an obscuration curve with different abundance of
silicates. Finally, some data at sub-mm wavelength would be needed
to better constrain the fit. Anyway, the results we have derived
suggest a population with an average age of about $100$ Myr. In the
case of this galaxy the SSP approximation is perhaps too crude and
it should be more appropriate to use of a mix of stellar
populations, thus perhaps getting a better fit of both the
absorption and emission features in the MIR.

\section{Discussion and conclusions}\label{dis_concl}

In this study we have presented detailed models of extinction and
emission of the diffuse ISM and the application of the most suitable
one to the calculation of a library of young dusty SSPs embedded in
their parental MCs.

We have analyzed the properties of a dusty ISM having as guide-line
of the strategy the cross-checking of the results for extinction and
emission and iterating the procedure to contrive the many, somehow
unavoidable parameters at work, so that not only unrealistic
solutions have been ruled out (they could by generated by the sole
fit of the extinction curve) but also additional information on the
whole problem has been acquired.

We have adopted two possible models for the distribution of the grains,
shortly named MRN and WEI, and explore the consequences. In brief,
the MRN-model is a multi-power-law, whose number of components
varies with the grain type, whereas the WEI-model is a complicated
analytical law suitably corrected for the smaller carbonaceous
grains by the sum of two log-normal terms. Again, we followed two
methods to derive the temperature distribution of VSGs. The first
method is a combination and modification of what has been proposed
from \citet{Guhathakurta89} for silicate and graphite grains and
from \citet{Puget85} for PAHs, in which we introduced the
calculation of the ionization state of PAHs following
\citet{Weingartner01b}. It has been shortly indicated as the
GDP-model. The second method, due to \citet{Draine01} and
\citet{Li01}, is the state-of-the-art of these problems as it stands
on the exact statistical solution of the problem. We have referred
to it as the LID model. Using the LID and/or GDP emission model
coupled with MRN and/or WEI model for the size distribution we have
calculated the extinction and emission properties of the average
diffuse ISM of the MW, LMC and SMC and compared it with
observational data.

Looking at the results from GDP+MRN and GDP+WEI and their comparison
with the LID+WEI, we adopt the GDP+WEI model as the most suitable
description of the properties of the dusty ISM to be used in dusty
SSPs and in the model of galaxy SEDs in presence of interstellar
dust \citep{Piovan05}.

We have  applied the model of dusty ISM to calculate the SEDs of
young SSPs, in which the effects of dust are taken into account.
These are particularly relevant in massive, young stars when still
embedded in their parental MCs. We propose a description in which
stars, gas and dust in a MC obey three different density profiles
each of these characterized by its own scale length. In this
context, we derive the optical depth of the MC and apply the ray
tracing method to solve the problem of radiative transfer across the
MC.
We have  also explored the parameter space of the MC models. Four
parameters have been taken into account, namely the scale radius of
the MC and its optical depth (the geometrical parameters), the
abundance of carbon in the log-normal populations of very small
carbonaceous grains and the ionization state of PAHs, both strongly
influencing the PAH emission. Basing on this, we have generated an
extended library of young dusty SSPs of different chemical
composition and age specifically taylored for galactic population
synthesis, the final goal of our study. We tested our SEDs trying to
reproduce the observational data of star forming regions in the
centre of local starburst galaxies, reaching a good agreement
between theory and observation.
Libraries of SSPs at varying these parameters are made available to
public for future use\footnote{Available from the authors upon
request on the web-page http://dipastro.pd.astro.it/galadriel}.

\section*{Acknowledgements}

We would like to deeply thank T. Takagi, A. Li, A. Weiss, B. T.
Draine, C. Joblin  for their explanations, for showing much interest
in our work, for the many clarifications and for kindly sending us
emission data. L.P. is pleased to acknowledge the hospitality and
stimulating environment provided by Max-Planck-Institut f\"ur
Astrophysik in Garching where a great deal of the work described in
this paper has been made during his visit as EARA fellow. This study
has been financed by the Italian Ministry of Education, University,
and Research (MIUR), and the University of Padua under the special
contract "Formation and evolution of elliptical galaxies: the age
problem".

\begin{small}
\bibliographystyle{mn2e}                    
\bibliography{mnemonic,Piovan2005}          
\end{small}
\end{document}